\documentclass[prx,twocolumn,superscriptaddress,nofootinbib,longbibliography]{revtex4-2}

\usepackage{amsmath,amssymb,bm}
\usepackage{graphicx}
\usepackage{xcolor}
\usepackage{booktabs}
\usepackage{quantikz}
\usetikzlibrary{decorations.pathmorphing,decorations.pathreplacing,arrows.meta,positioning,calc}
\usepackage[colorlinks=true,linkcolor=blue,citecolor=blue,urlcolor=blue]{hyperref}

\providecommand{\ket}[1]{\lvert #1 \rangle}
\providecommand{\bra}[1]{\langle #1 \rvert}
\providecommand{\ketbra}[2]{\lvert #1 \rangle\!\langle #2 \rvert}
\newcommand{\dt}{\Delta t}
\newcommand{\Tr}{\operatorname{Tr}}
\newcommand{\D}{\mathcal{D}}
\newcommand{\Hel}{H_{\mathrm{S}}}
\newcommand{\Hdil}{H_{\mathrm{dil}}}
\newcommand{\Udil}{U_{\mathrm{dil}}}
\newcommand{\nhat}{\hat{n}}
\newcommand{\CX}{\textsc{cx}}

\begin{document}
\title{Quantum simulation of non-Markovian dynamics with coupled Lindblad pseudomodes: Exact momentum-space factorization and fault-tolerant resources}

\author{Mauro Cainelli}
\email{cainelli@theoc.kuchem.kyoto-u.ac.jp}
\affiliation{Department of Chemistry, Graduate School of Science, Kyoto University, Kitashirakawa Oiwake-cho, Sakyo-ku, Kyoto 606-8502, Japan}

\author{Yuki Kurashige}
\email{kura@kuchem.kyoto-u.ac.jp}
\affiliation{Department of Chemistry, Graduate School of Science, Kyoto University, Kitashirakawa Oiwake-cho, Sakyo-ku, Kyoto 606-8502, Japan}
\affiliation{FOREST, JST, Honcho 4-1-8, Kawaguchi, Saitama 332-0012, Japan}
\affiliation{CREST, JST, Honcho 4-1-8, Kawaguchi, Saitama 332-0012, Japan}

\begin{abstract}
Non-Markovian open-system dynamics can be cast in Lindblad form by embedding the memory of the environment in a small set of damped bosonic modes, known as pseudomodes. 
When the pseudomodes are allowed to couple to each other, the resulting completely positive generator reproduces correlation functions with complex amplitudes that independent Lorentzian modes cannot, and
its dynamics unravels into quantum trajectories that can be simulated on a quantum computer with a single reusable ancilla and no postselection. Here we construct and benchmark quantum-trajectory circuits for coupled bosonic pseudomodes and estimate their fault-tolerant cost. Validated against the hierarchical equations of motion for Drude--Lorentz and Brownian baths, two coupled modes per bath reproduce the exact dynamics within the accuracy of the fit. We introduce two transformations: an encoding in the eigenmodes of the pseudomode Hamiltonian, which moves the inter-mode coupling into the dissipator, and, for $N$ sites with identical baths, a discrete Fourier transform over the bath index that removes one bath exactly and turns the remaining couplings into phase waves whose propagators factorize exactly, at a cost that grows as $\log N$ per block.
Together with further compilation steps, these transformations reduce the cost of a Trotter step by two orders of magnitude and make it grow nearly linearly with the number of baths. A source-by-source error budget shows that all numerical approximations remain several times below the error of the pseudomode fit, and that the Stinespring dilation requires at least a second-order product formula for its implementation. The resulting circuits require few logical qubits, and their end-to-end cost is set by the number of time steps and trajectories rather than by compilation.
\end{abstract}

\maketitle

\section{Introduction}
\label{sec:intro}

Charge and energy transfer in molecular aggregates and photosynthetic complexes, vibronic relaxation in condensed-phase spectroscopy, and decoherence in solid-state quantum devices are all processes in which a small system interacts with a large environment~\cite{Breuer2002,Weiss2012,Leggett1987,Berkelbach2020}. In many of these processes, the environment is structured: its spectral density has sharp features, the system--bath coupling is not weak, and its correlation time is comparable to the system timescales. The reduced dynamics then retains memory of its past, and information flows back from the environment into the system~\cite{deVega2017}. Simulating such non-Markovian dynamics is a central task of open-system theory, and because the composite system--bath state lives in an exponentially large Hilbert space, it is also a natural target for quantum computers~\cite{Feynman1982,Lloyd1996,mcardle2020yuan,bauer2020chan}.

When the bath correlations decay fast compared with the system dynamics and the coupling is weak, the reduced density matrix obeys a Lindblad master equation~\cite{Gorini1976,Lindblad1976}, whose generator is completely positive and trace preserving (CPTP) by construction. A broad family of quantum algorithms simulates Lindbladians, from product formulas and Kraus-operator dilations~\cite{Kliesch2011,Childs2017,Hu2020,Ding2024} and unitary decompositions of the non-unitary propagator~\cite{schlimgen2021mazziotti,schlimgen2022mazziotti,burdine2024blair} to quantum imaginary-time evolution~\cite{kamakari2022minnich-1}, linear combinations of unitaries and higher-order series expansions~\cite{Cleve2017,Li2023}, randomized dissipation~\cite{Kato2024}, and trajectory-inspired channels~\cite{Peng2025}.
The simplest of them follows the stochastic unraveling of the master equation into quantum trajectories~\cite{Dalibard1992,Plenio1998}, which can be implemented on a quantum device by dilating each jump into a unitary operator acting on the system and one ancilla qubit~\cite{Stinespring1955}. The ancilla is then measured and reset after every channel, and averaging over repeated circuit executions (shots) reproduces the density matrix~\cite{Ding2024}. Both measurement outcomes contribute to the dynamics and are kept, so no postselection is required. The price of this simplicity is that the generator must be of Lindblad form, with a Hermitian Hamiltonian and a positive dissipator.

Non-Markovian dynamics do not satisfy this requirement. The hierarchical equations of motion (HEOM)~\cite{Tanimura1989,Tanimura2020} provide a numerically exact treatment of Gaussian baths by expanding the bath correlation function (BCF) in exponentials and propagating a hierarchy of auxiliary density operators (ADOs) that store the memory. The resulting amplitudes are complex, and the HEOM generator is neither Hermitian nor CPTP: its bra and ket components evolve differently, and the hierarchy must be propagated in Liouville space. Quantum implementations of HEOM are possible, but they require doubling the register to represent the density operator and dilating non-unitary propagators with postselection~\cite{Dan2025}. General non-unitary propagators can also be expressed as linear combinations of Hamiltonian simulations~\cite{an2026lin}. Other approaches encode the exact non-Markovian dynamics in Kraus operators implemented through unitary dilations~\cite{seneviratne2024wang}, or exploit the intrinsic dissipation of near-term hardware to emulate structured vibrational environments in digital-analog simulations~\cite{gajewski2026horstmann}. Classically, numerically exact methods already reach systems of considerable size: HEOM combined with tensor-network representations of the hierarchy~\cite{Borrelli2021,Takahashi2024,ke2022thoss}, chain-mapping methods such as TEDOPA~\cite{Prior2010,Tamascelli2019}, the time-evolving matrix product operator method~\cite{Strathearn2018}, multilayer wave-packet methods~\cite{Wang2003}, positivity-preserving tensor networks for dissipative dynamics~\cite{werner2016montangero}, and dissipation-assisted matrix product factorization (DAMPF)~\cite{Somoza2019,Lorenzoni2024}, which has recently simulated the complete Fenna--Matthews--Olson complex with structured spectral densities~\cite{Lorenzoni2025}.

Pseudomode theory~\cite{Garraway1997,Dalton2001,Tamascelli2018} offers an alternative route. The continuous bath is replaced by a few damped bosonic modes whose own correlation function reproduces the BCF; because the reduced dynamics of a Gaussian bath depends only on its BCF, the system dynamics is unchanged. When each mode is independent and damped by a Lindblad dissipator (Lorentzian pseudomodes), the joint dynamics is CPTP, and each mode contributes a term with a real, positive weight to the BCF. Such modes can approximate a finite-temperature BCF, as in DAMPF~\cite{Somoza2019,Mascherpa2017}, but their number grows polynomially with the simulated time and precision~\cite{Trivedi2021}, and they cannot reproduce the Matsubara or Pad\'e terms of HEOM decompositions~\cite{Hu2010}, which carry real rates and negative weights, so that the low-temperature corrections require many modes. Proposed solutions, including unphysical modes~\cite{Lambert2019,Cirio2024}, non-Hermitian pseudomodes~\cite{Menczel2024}, and quasi-Lindblad pseudomodes~\cite{Park2024}, recover efficiency by giving up complete positivity, and with it the direct unraveling into trajectories on quantum hardware.

Coupled Lindblad pseudomodes~\cite{Dorda2014,Dorda2015,Mascherpa2020,Pleasance2020,Lednev2024,Huang2026,Muller2026} resolve this tension. The mode Hamiltonian and the dissipation matrix are allowed to be dense Hermitian matrices, which keeps the generator CPTP, while their off-diagonal elements reproduce complex exponentials with complex weights. Huang \emph{et al.}~\cite{Huang2026} showed that the number of coupled modes needed for a precision $\varepsilon$ over a time $T$ scales as $\mathrm{polylog}(T/\varepsilon)$, and introduced a convex, seed-free construction based on the realization problem of control theory. M\"uller and Strunz~\cite{Muller2026} proved that every physical BCF consisting of $N$ exponentials can be represented exactly by $N$ coupled pseudomodes, only one of which needs to be damped, and established a one-to-one correspondence between the pseudomode and HEOM descriptions. Coupled pseudomodes are therefore the completely positive counterpart of HEOM: they combine the compactness of an optimized exponential fit with exactly the property that quantum trajectory algorithms need.

Pseudomode-based and related approaches have begun to reach quantum circuits: a collision-model algorithm with Lorentzian pseudomodes for multidimensional electronic spectroscopy~\cite{Spectroscopy2024}, an ancilla-based algorithm for general Gaussian environments with rigorous error bounds~\cite{AncillaTrain2025}, and Stinespring dilations of a two-level pseudomode for open dynamics in heavy-ion collisions~\cite{HeavyIon2026}. On native bosonic hardware, pseudomodes can be realized directly as damped motional modes of trapped ions~\cite{Lemmer2018}, as in a recent simulation of electron-transfer models of the same class~\cite{So2024}. What has been missing, to our knowledge, is a circuit implementation of coupled pseudomodes for relevant systems, its validation against exact methods such as HEOM, and a quantitative account of the cost of compiling it into gate-based fault-tolerant circuits: how the gate count depends on the bosonic encoding and on the number of system sites, which part of the circuit dominates the numerical error, and how the result compares with general-purpose Lindbladian algorithms.

In this work we assess whether simulations along this path are feasible on future fault-tolerant quantum computers. As reference we use classically tractable systems: the spin-boson model, and excitonic dimers and trimers with the parameters of perylene bisimide (PBI) aggregates~\cite{Cainelli2022}, in which electronic sites couple to underdamped Brownian baths. Our main contributions are the following.
(i) We construct quantum-trajectory circuits for coupled pseudomodes fitted to Drude--Lorentz and Brownian spectral densities, based on Stinespring dilations with a single reusable ancilla~\cite{Ding2024}, and validate them against HEOM.
(ii) We introduce a transformation to the eigenmodes of the pseudomode Hamiltonian, which moves the inter-mode coupling into the dissipator and drastically reduces the cost of the coherent part of the circuit.
(iii) For $N$ sites with identical and independent baths, we introduce a discrete Fourier transform over the bath index that decouples one bath exactly and maps the remaining $N-1$ baths onto phase-wave couplings, whose propagators factorize exactly into two diagonal phase layers and a single-mode unitary, with a cost that grows as $\log N$ per block.
(iv) We isolate the contribution of each approximation to the numerical error, estimate the fault-tolerant resources and their scaling with system size, and compare the approach with other quantum and classical methods.

The paper is organized as follows. Sec.~\ref{sec:model} reviews HEOM, pseudomode and coupled-pseudomode theory, and the fitting of the bath correlation function. Sec.~\ref{sec:circuit} describes the quantum-trajectory circuit and the accuracy of the dilated channel, and Sec.~\ref{sec:compilation} the structure-exploiting transformations. Sec.~\ref{sec:results} presents the results: the validation on a single shared bath with Drude--Lorentz and Brownian spectral densities, the dynamics of chains with independent baths together with the error budget and the fault-tolerant resources, and a comparison with other quantum and classical approaches. Sec.~\ref{sec:conclusions} summarizes our conclusions.

\section{Theoretical framework}
\label{sec:model}

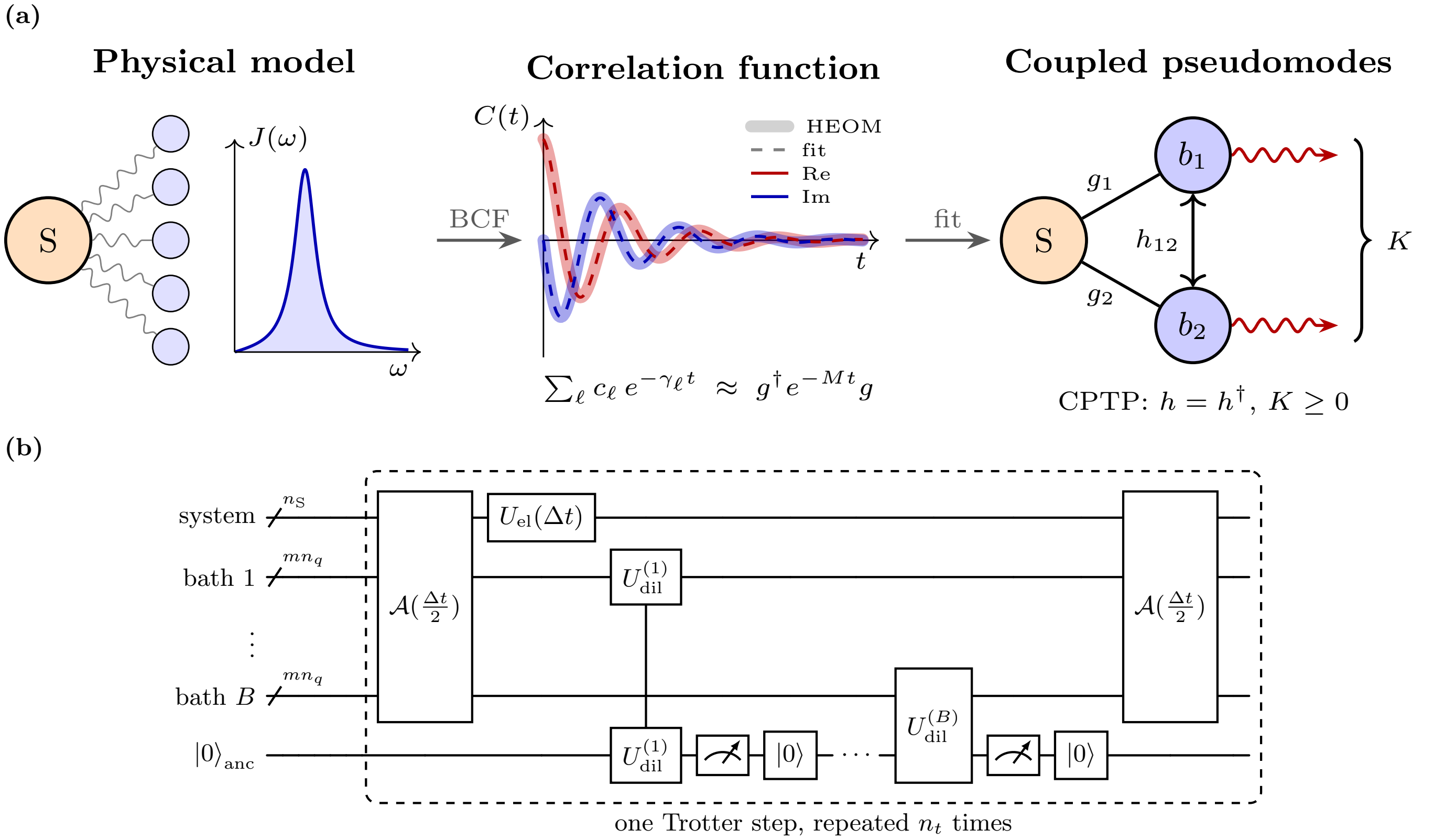
\begin{figure*}[t]
{\raggedright\textbf{(a)}\par}
\vspace{2pt}
\centering
\resizebox{\textwidth}{!}{%
\begin{tikzpicture}[font=\small,
  stage/.style={font=\small\bfseries, anchor=south},
  flow/.style={-{Stealth[length=2.2mm]}, thick, gray!70!black},
  sysnode/.style={circle, draw, thick, fill=orange!25, minimum size=8mm, inner sep=0pt},
  osc/.style={circle, draw, fill=blue!12, minimum size=3.4mm, inner sep=0pt},
  pm/.style={circle, draw, thick, fill=blue!20, minimum size=7mm, inner sep=0pt},
  coup/.style={decorate, decoration={snake, amplitude=0.6mm, segment length=2.2mm}},
  damp/.style={decorate, decoration={snake, amplitude=0.6mm, segment length=2.2mm, post length=1.5mm}, -{Stealth[length=1.8mm]}, thick, red!70!black},
  band/.style={line width=3.2pt, opacity=0.35, line cap=round},
]
\node[stage] at (1.95,2.6) {Physical model};
\node[sysnode] (S1) at (0.3,1.2) {S};
\foreach \y [count=\i] in {2.2,1.7,1.2,0.7,0.2} {
  \node[osc] (o\i) at (1.45,\y) {};
  \draw[coup, gray] (S1) -- (o\i);
}
\fill[blue!12, domain=0.01:6.8, samples=200, smooth]
  (2.05,0.15) -- plot ({2.05+0.24*\x}, {0.15+1.7*(2*2.4*7.84*\x/((7.84-\x*\x)^2+\x*\x))/13.44}) -- (3.68,0.15) -- cycle;
\draw[thick, blue!70!black, domain=0.01:6.8, samples=200, smooth]
  plot ({2.05+0.24*\x}, {0.15+1.7*(2*2.4*7.84*\x/((7.84-\x*\x)^2+\x*\x))/13.44});
\draw[->] (2.05,0.15) -- (3.8,0.15) node[below left, font=\scriptsize] {$\omega$};
\draw[->] (2.05,0.15) -- (2.05,2.15) node[right, font=\scriptsize] {$J(\omega)$};
\draw[flow] (3.95,1.2) -- node[above, font=\scriptsize] {BCF} (4.75,1.2);
\begin{scope}[shift={(4.95,0.1)}]
  \node[stage] at (1.5,2.5) {Correlation function};
  \draw[->] (0,1.1) -- (3.15,1.1) node[below left] {\scriptsize $t$};
  \draw[->] (0,0) -- (0,2.25) node[left] {\scriptsize $C(t)$};
  \draw[band, red!80!black, domain=0:3.0, samples=150, smooth]
    plot (\x, {1.1+0.95*exp(-1.6*\x)*cos(deg(8.5*\x))});
  \draw[band, blue!80!black, domain=0:3.0, samples=150, smooth]
    plot (\x, {1.1-0.95*exp(-1.6*\x)*sin(deg(8.5*\x))});
  \draw[thick, dashed, red!70!black, domain=0:3.0, samples=150, smooth]
    plot (\x, {1.1+0.95*exp(-1.6*\x)*cos(deg(8.5*\x))});
  \draw[thick, dashed, blue!70!black, domain=0:3.0, samples=150, smooth]
    plot (\x, {1.1-0.95*exp(-1.6*\x)*sin(deg(8.5*\x))});
  \draw[band, gray] (1.95,2.17) -- (2.3,2.17) node[right, font=\tiny, black, opacity=1] {HEOM};
  \draw[thick, dashed, gray] (1.95,1.95) -- (2.3,1.95) node[right, font=\tiny, black] {fit};
  \draw[thick, red!70!black] (1.95,1.73) -- (2.3,1.73) node[right, font=\tiny, black] {Re};
  \draw[thick, blue!70!black] (1.95,1.51) -- (2.3,1.51) node[right, font=\tiny, black] {Im};
  \node[font=\scriptsize] at (1.55,-0.28)
    {$\sum_\ell c_\ell\, e^{-\gamma_\ell t}\;\approx\; g^\dagger e^{-Mt} g$};
\end{scope}
\draw[flow] (8.35,1.2) -- node[above, font=\scriptsize] {fit} (9.15,1.2);
\begin{scope}[shift={(9.35,0)}]
  \node[stage] at (1.75,2.6) {Coupled pseudomodes};
  \node[sysnode] (S3) at (0.3,1.2) {S};
  \node[pm] (b1) at (1.7,2.0) {$b_1$};
  \node[pm] (b2) at (1.7,0.4) {$b_2$};
  \draw[thick] (S3) -- node[above left, font=\scriptsize, inner sep=1pt] {$g_1$} (b1);
  \draw[thick] (S3) -- node[below left, font=\scriptsize, inner sep=1pt] {$g_2$} (b2);
  \draw[thick, <->] (b1) -- node[left, font=\scriptsize] {$h_{12}$} (b2);
  \draw[damp] (b1.east) -- ++(1.0,0) coordinate (e1);
  \draw[damp] (b2.east) -- ++(1.0,0) coordinate (e2);
  \draw[decorate, decoration={brace, amplitude=4pt}, thick] ($(e1)+(0.15,0.15)$) -- ($(e2)+(0.15,-0.15)$)
    node[midway, right=5pt, font=\scriptsize] {$K$};
  \node[font=\scriptsize] at (1.8,-0.3) {CPTP: $h=h^\dagger$, $K\geq 0$};
\end{scope}
\end{tikzpicture}}

\vspace{0pt}
{\raggedright\textbf{(b)}\par}
\vspace{2pt}
\centering
\begin{quantikz}[column sep=0.2cm, row sep={0.75cm,between origins}]
\lstick{system} & \qwbundle{n_{\mathrm{S}}} & & & & & & \gate[4]{\mathcal{A}(\tfrac{\dt}{2})}\gategroup[5,steps=12,style={dashed,rounded corners,inner xsep=1pt},background,label style={label position=below,anchor=north,yshift=-0.2cm}]{one Trotter step, repeated $n_t$ times} & \gate{U_{\mathrm{el}}(\dt)} & & & & & & & & \gate[4]{\mathcal{A}(\tfrac{\dt}{2})} & & \\
\lstick{bath $1$} & \qwbundle{m n_q} & & & & & & & & \gate{U_{\mathrm{dil}}^{(1)}}\vqw{3} & & & & & & & & & \\
\lstick{$\vdots$} \setwiretype{n} & & & & & & & & & & & & & & & & & & \\
\lstick{bath $B$} & \qwbundle{m n_q} & & & & & & & & & & & & \gate[2]{U_{\mathrm{dil}}^{(B)}} & & & & & \\
\lstick{$\ket{0}_{\mathrm{anc}}$} & & & & & & & & & \gate{U_{\mathrm{dil}}^{(1)}} & \meter{} & \gate{\ket{0}} & \ \cdots\ & & \meter{} & \gate{\ket{0}} & & &
\end{quantikz}
\caption{(a) Overview of the method. A system S coupled to a harmonic bath with spectral density $J(\omega)$ is characterized by its bath correlation function $C(t)$. Its exponential decomposition, used by HEOM (thick lines, real and imaginary parts), is fitted by the correlation function $g^\dagger e^{-Mt}g$ of a few coupled pseudomodes (dashed lines). The pseudomodes interact coherently through the Hermitian matrix $h$ and are damped through the positive semidefinite dissipation matrix $K$, which can also couple them. (b) Quantum-trajectory circuit, in which the system and the pseudomodes are encoded in qubits. Each Trotter step follows the symmetric sequence of Sec.~\ref{sec:circuit:register}, with the coherent bath block $\mathcal{A}$, the system propagator $U_{\mathrm{el}}$, and one dissipative channel $\mathcal{C}$ per bath $b=1,\dots,B$ and jump operator. Each channel consists of the dilation unitary $U_{\mathrm{dil}}^{(b)}$ acting on bath $b$ and the ancilla, followed by measurement and reset of the ancilla; a single ancilla is reused by all channels, and both outcomes are kept. Each execution of the circuit yields one trajectory.}
\label{fig:circuit}
\end{figure*}

\subsection{System--bath model and bath correlation functions}
\label{sec:model:sb}

We consider a system coupled linearly to a harmonic environment,
\begin{equation}
  H_{\mathrm{tot}} = \Hel + \sum_{\alpha} V_\alpha \otimes X_\alpha + H_{\mathrm{B}},
  \label{eq:Htot}
\end{equation}
where $V_\alpha$ is a Hermitian system operator, $X_\alpha$ a linear combination of the coordinates of bath $\alpha$, and $H_{\mathrm{B}}$ a set of harmonic oscillators, initially in thermal equilibrium at inverse temperature $\beta$, so that the initial state is
$\rho_{\mathrm{S}}(0)\otimes\rho_{\mathrm{B}}^{\mathrm{eq}}$.
For a Gaussian environment, its influence on the system is fully determined by the bath correlation functions (BCFs),
\begin{equation}
  C_{\alpha\beta}(t) = \Tr_{\mathrm{B}}\!\left[X_\alpha(t)X_\beta(0)\,\rho_{\mathrm{B}}^{\mathrm{eq}}\right],
  \label{eq:Cab}
\end{equation}
where $X_\alpha(t)$ evolves under $H_{\mathrm{B}}$. These are usually expressed in terms of the spectral densities $J_{\alpha\beta}(\omega)$
\begin{align}
  C_{\alpha\beta}(t) = \frac{1}{\pi}\int_0^\infty d\omega\, J_{\alpha\beta}(\omega)
  &\left[\coth\!\left(\frac{\beta\omega}{2}\right)\cos\omega t\right. \nonumber\\
  &\left. \vphantom{\coth\!\left(\frac{\beta\omega}{2}\right)} - i\sin\omega t\right],
  \label{eq:Ct}
\end{align}
where the BCFs satisfy $C_{\alpha\beta}(t)^{*}=C_{\beta\alpha}(-t)$. Two models with the same BCF matrix produce the same reduced dynamics of the system~\cite{Tamascelli2018}. This equivalence underlies both methods discussed below: HEOM encodes the BCFs in a hierarchy of auxiliary operators, and pseudomode theory replaces the environment by a few damped physical modes that reproduce the same BCFs.

\subsection{Hierarchical equations of motion}
\label{sec:model:heom}

HEOM~\cite{Tanimura1989,Tanimura2020} starts from an expansion of the BCFs in exponentials,
\begin{equation}
  C_{\alpha\beta}(t) \simeq \sum_{\ell=1}^{N_{\mathrm{exp}}} c^{\alpha\beta}_\ell\, e^{-\gamma_\ell t}, \qquad
  C_{\alpha\beta}(t)^{*} \simeq \sum_{\ell=1}^{N_{\mathrm{exp}}} \bar c^{\,\alpha\beta}_\ell\, e^{-\gamma_\ell t},
  \label{eq:Cexp}
\end{equation}
for $t\geq0$. Depending on the spectral density, the amplitudes $c^{\alpha\beta}_\ell$ and the rates $\gamma_\ell$ ($\mathrm{Re}\,\gamma_\ell>0$) are in general complex. They can be obtained analytically for some spectral densities, in which case the peaks of $J_{\alpha\beta}(\omega)$ give damped oscillatory or purely decaying terms and the poles of $\coth(\beta\omega/2)$ give decaying low-temperature corrections (Matsubara or Pad\'e terms~\cite{Hu2010}), or numerically by fitting the BCF~\cite{Xu2022,xu2023ankerhold,TakahashiFit2024}.

Assigning one index $n_{\alpha\ell}\geq0$ to each channel and exponential defines a hierarchy of auxiliary density operators (ADOs) $\rho_{\mathbf n}$, whose zeroth member $\rho_{\mathbf 0}$ is the reduced density matrix of the system. The resulting HEOM can be written as~\cite{Tanimura2020}
\begin{align}
  \dot\rho_{\mathbf n} &= -\Big(i\Hel^{\times} + \sum_{\alpha\ell} n_{\alpha\ell}\gamma_\ell\Big)\rho_{\mathbf n}
  - i\sum_{\alpha\ell}V_\alpha^{\times}\rho_{\mathbf n^{+}_{\alpha\ell}} \nonumber\\
  &\quad - \sum_{\alpha\ell} n_{\alpha\ell}\sum_{\beta}\left(i\,s^{\alpha\beta}_\ell V_\beta^{\times} - a^{\alpha\beta}_\ell V_\beta^{\circ}\right)\rho_{\mathbf n^{-}_{\alpha\ell}},
  \label{eq:heom}
\end{align}
with $A^{\times}\rho=[A,\rho]$ and $A^{\circ}\rho=\{A,\rho\}$. Here, $n_{\alpha\ell}$ is the hierarchy index, which is raised or lowered by one, and the hierarchy is truncated at a maximum tier $\sum_{\alpha\ell}n_{\alpha\ell}\leq\mathcal{N}_{\max}$. The coefficients $s^{\alpha\beta}_\ell$ and $a^{\alpha\beta}_\ell$ are the amplitudes of the symmetrized and antisymmetrized correlation functions~\cite{Ikeda2020},
\begin{align}
  \tfrac{1}{2}\big\langle\{X_\alpha(t),X_\beta(0)\}\big\rangle &= \mathrm{Re}\,C_{\alpha\beta}(t) \simeq \sum_\ell s^{\alpha\beta}_\ell e^{-\gamma_\ell t},
  \label{eq:Sexp}\\
  \tfrac{1}{2i}\big\langle[X_\alpha(t),X_\beta(0)]\big\rangle &= \mathrm{Im}\,C_{\alpha\beta}(t) \simeq \sum_\ell a^{\alpha\beta}_\ell e^{-\gamma_\ell t},
  \label{eq:Aexp}
\end{align}
which describe the fluctuation and the dissipation of the environment, respectively, and are related to the amplitudes of Eq.~\eqref{eq:Cexp} by $c^{\alpha\beta}_\ell=s^{\alpha\beta}_\ell+i\,a^{\alpha\beta}_\ell$ and $\bar c^{\,\alpha\beta}_\ell=s^{\alpha\beta}_\ell-i\,a^{\alpha\beta}_\ell$.

The ADOs store the memory of the environment: the second term of Eq.~\eqref{eq:heom} couples each tier to the next through the system--bath interaction, and the last term feeds back fluctuation and dissipation from the tier below. For uncorrelated channels, $c^{\alpha\beta}_\ell=\delta_{\alpha\beta}c^{\alpha}_\ell$, the sum over $\beta$ reduces to the single term $\beta=\alpha$. With the exponential expansion converged and $\mathcal{N}_{\max}$ large enough, Eq.~\eqref{eq:heom} is numerically exact, and we use it as the reference throughout.

Equation~\eqref{eq:heom} is linear in the ADOs but is not a Lindblad equation. The amplitudes $c_\ell$ and $\bar c_\ell$ differ and are complex, so the bra and ket sides of each ADO evolve differently, the ADOs are not Hermitian, and the hierarchy generator is not CPTP. Moreover, the target BCF need not be regular at $t=0$. For spectral densities that decay only as $1/\omega$, such as the Drude--Lorentz form, $\mathrm{Im}\,C_{\alpha\alpha}(0^{+})\neq0$ and $\mathrm{Re}\,C_{\alpha\alpha}(0)$ diverges, so that the BCF is discontinuous at $t=0$; truncated Matsubara or Pad\'e expansions introduce further errors at short times. Classically this is harmless as long as the propagation is stable, which is not guaranteed in general~\cite{Dunn2019,yan2020shi,Krug2023}. On a quantum computer, however, this forces the hierarchy to be represented in Liouville space, doubling the register, and its non-unitary propagator to be implemented by dilation with postselection~\cite{Dan2025}.

\subsection{Pseudomode representation}
\label{sec:model:lorentz}

Pseudomode theory~\cite{Garraway1997,Tamascelli2018} replaces the environment by a finite set of bosonic modes $b_j$ ($j=1,\dots,m$), initially in the vacuum and damped by a Lindblad dissipator $\D$, chosen so that their correlation functions equal the BCFs of the original environment. In the simplest construction, the modes are independent and the density matrix evolution can be expressed as
\begin{align}
  \dot\rho &= -i[H,\rho] + \sum_{j}\kappa_j\,\D[b_j]\rho, \label{eq:pm}\\
  H &= \Hel + \sum_{\alpha}V_\alpha\otimes\sum_j\left(g_{\alpha j}^{*}b_j + g_{\alpha j} b_j^\dagger\right) \nonumber\\
  &\quad + \sum_{j}\omega_j\, b_j^\dagger b_j, \nonumber
\end{align}
with complex couplings $g_{\alpha j}$ and the Lindblad dissipator
$\D[O]\rho = O\rho O^\dagger - \tfrac12\{O^\dagger O,\rho\}$, defined for any operator $O$. Here, the jump operators are $\sqrt{\kappa_j}\,b_j$, where each mode is damped independently at rate $\kappa_j$. The joint dynamics is CPTP, and each mode contributes one exponential to the BCFs
\begin{equation}
  C^{\mathrm{PM}}_{\alpha\beta}(t) = \sum_j g_{\alpha j}^{*}g_{\beta j}\, e^{-(\kappa_j/2 + i\omega_j)t},
  \label{eq:Clorentz}
\end{equation}
corresponding to a Lorentzian peak of the spectral densities. The amplitude matrix of each exponential, $g_{\alpha j}^{*}g_{\beta j}$, is positive semidefinite; in particular, the amplitudes of the autocorrelation functions are real and positive. The low-temperature terms of Eq.~\eqref{eq:Cexp}, and in general any decomposition with complex or negative amplitudes, can therefore not be reproduced by independent modes. Allowing non-Hermitian couplings or mode Hamiltonians~\cite{Lambert2019,Cirio2024,Menczel2024}, or relaxing the Lindblad form of the dissipator~\cite{Park2024}, reproduces such BCFs efficiently, but breaks complete positivity. Such generators can still be unraveled into trajectories, but only with averages weighted by a probability pseudo-measure~\cite{donvil2022muratore-ginanneschi}, or mapped onto Lindblad equations in an enlarged space~\cite{hush2015garrahan}. On quantum circuits, non-CPTP generators, like HEOM itself, can still be simulated on quantum circuits by representing the density operator in a doubled ket--bra register and implementing the anti-Hermitian part of the generator with probabilistic imaginary-time evolution (PITE) gadgets~\cite{Liu2021,Kosugi2022}. We implemented this approach and verified that it reproduces the HEOM dynamics; since a detailed analysis is beyond the scope of the present work, we only summarize its limitations here. Every gadget must be postselected, so the success probability decays exponentially with the simulated time independently of the time step; the normalization must be tracked through weights that depend on the success probabilities; and the populations are encoded in amplitudes rather than in measurement probabilities, which requires interferometric readout. Coupled Lindblad pseudomodes avoid all three requirements.

\subsection{Coupled Lindblad pseudomodes}
\label{sec:model:pm}

Coupled pseudomodes~\cite{Mascherpa2020,Pleasance2020,Lednev2024,Huang2026,Muller2026} keep the Lindblad form but allow the modes to interact through both the Hamiltonian and the dissipator,
\begin{align}
  \dot\rho &= -i[H,\rho] + \sum_{ij} K_{ij}\left(b_j\rho\, b_i^\dagger - \tfrac{1}{2}\{b_i^\dagger b_j,\rho\}\right),
  \label{eq:clpm}\\
  H &= \Hel + \sum_{\alpha}V_\alpha\otimes\sum_j\left(g_{\alpha j}^{*}\, b_j + g_{\alpha j}\, b_j^\dagger\right) \nonumber\\
  &\quad + \sum_{ij} h_{ij}\, b_i^\dagger b_j,
  \nonumber
\end{align}
where $g_\alpha\in\mathbb{C}^m$ is the coupling vector of channel $\alpha$, $h=h^\dagger$ is the Hermitian mode Hamiltonian, and $K=K^\dagger\geq0$ is the positive semidefinite dissipation matrix. These conditions guarantee that Eq.~\eqref{eq:clpm} is of Lindblad form and therefore generates a CPTP map. Diagonalizing $K=U\,\mathrm{diag}(\kappa)\,U^\dagger$ makes this explicit,
\begin{align}
  \sum_{ij} K_{ij}\left(b_j\rho\, b_i^\dagger - \tfrac{1}{2}\{b_i^\dagger b_j,\rho\}\right)
  &= \sum_{\nu}\D[L_{\nu}]\rho, \label{eq:jumps}\\
  L_{\nu} = \sqrt{\kappa_\nu}\sum_j U^{*}_{j\nu}\, b_{j}, \nonumber
\end{align}
with one dissipative channel per nonzero eigenvalue $\kappa_\nu$. The jump operators $L_\nu$ are linear combinations of the annihilation operators of all modes, thus, for diagonal $h$ and $K$ they reduce to $\sqrt{\kappa_j}\,b_j$ and Eq.~\eqref{eq:pm} is recovered.

Since the dynamics of the modes is quadratic and they start in the vacuum, the Heisenberg equations close on the annihilation operators, $\dot b = -Mb$ with
\begin{equation}
  M = \frac{K}{2} + i h,
  \label{eq:M}
\end{equation}
and the correlation functions of the bath operators $X^{\mathrm{PM}}_\alpha=\sum_j (g_{\alpha j}^{*}b_j + g_{\alpha j} b_j^\dagger)$ are
\begin{equation}
  C^{\mathrm{PM}}_{\alpha\beta}(t) = g_\alpha^\dagger\, e^{-Mt}\, g_\beta .
  \label{eq:Cpm}
\end{equation}
When $C^{\mathrm{PM}}_{\alpha\beta}=C_{\alpha\beta}$ for all $\alpha,\beta$, the reduced system dynamics generated by Eq.~\eqref{eq:clpm} coincides with that of the original model~\cite{Tamascelli2018,Huang2026}. Diagonalizing $M$ shows that Eq.~\eqref{eq:Cpm} is a sum of $m$ complex exponentials whose rates are the eigenvalues of $M$. For independent modes, the amplitude matrices are those of Eq.~\eqref{eq:Clorentz}. For coupled modes, $M$ is in general non-normal, $[M,M^\dagger]\neq 0$, and the amplitudes become complex, which is what allows coupled modes to reproduce the complex amplitudes of Eq.~\eqref{eq:Cexp} while preserving the Lindblad form. No unitary change of mode basis diagonalizes $h$ and $K$ simultaneously; but a basis change allows one to choose whether the inter-mode coupling appears in the Hamiltonian or in the dissipator, a freedom we exploit in Sec.~\ref{sec:compilation:nm}.

The pseudomodes start in the vacuum at every temperature, its effect enters only through the target BCFs, and the thermal factor of Eq.~\eqref{eq:Ct} is absorbed into $g_\alpha$, $h$ and $K$. In the numerical representation, each mode is truncated to its lowest $d$ Fock states and encoded in $n_q=\log_2 d$ qubits; the Fock levels play a role analogous to the tiers of the hierarchy in HEOM.

Particularly, if the channels couple to independent baths, $C_{\alpha\beta}=\delta_{\alpha\beta}C_\alpha$, the modes can be grouped by bath, $b_j\to b_{\alpha j}$, and $h$ and $K$ become block diagonal, with each $g_\alpha$ supported on the modes of its own bath. For identical baths, all blocks share the same $(g,h,K)$ of dimension $m$, so that $C_\alpha(t)=g^\dagger e^{-Mt}g$ for every $\alpha$ and the jump operators of bath $\alpha$ are $L_{\alpha\nu}=\sqrt{\kappa_\nu}\sum_j U^{*}_{j\nu}\, b_{\alpha j}$. This is the notation used in Secs.~\ref{sec:circuit} and~\ref{sec:compilation}.

\subsection{Fitting the correlation functions}
\label{sec:model:fit}

The parameters $(g_\alpha,h,K)$ are determined by nonlinear least-squares minimization of $\sum_{t}\sum_{\alpha\beta}|C^{\mathrm{PM}}_{\alpha\beta}(t)-C_{\alpha\beta}(t)|^2$ over a uniform time grid, in which the real and imaginary parts of the residual enter with equal weight and $C^{\mathrm{PM}}$ is evaluated from Eq.~\eqref{eq:Cpm} by matrix exponentiation with the constraints built into the parametrization. The dissipation matrix is written as $K=RR^\dagger$, with $R$ lower triangular, complex off-diagonal elements, and diagonal elements $R_{jj}=e^{x_j}>0$, which guarantees $K\geq0$; $h$ is parametrized by its real diagonal and complex lower triangle, and $g_\alpha$ by its real and imaginary parts. 
The minimization uses the trust-region reflective algorithm of SciPy's \texttt{least\_squares} with finite-difference Jacobians, starting from $K\propto\mathbb{1}$, $h=0$, and a random $g$, and is repeated over several random initializations to locate the global optimum. For independent baths, the fit separates into one scalar fit per distinct BCF, and for identical baths, into a single one. The fitted parameters are not unique: transformations of the modes that leave Eq.~\eqref{eq:Cpm} and the constraints invariant, such as a unitary change of mode basis, yield equivalent parameter sets, and the reduced dynamics depends only on $C^{\mathrm{PM}}$. The target can be the exact BCFs of Eq.~\eqref{eq:Ct} or the exponential decomposition used by HEOM, Eq.~\eqref{eq:Cexp}. Fitting the latter makes the pseudomode and HEOM calculations describe the same environment, so that the difference between them isolates the error of the fit. If the target is a physical BCF consisting of $N$ exponentials, an exact representation with $N$ coupled pseudomodes exists~\cite{Muller2026}, and the number of exponentials needed for precision $\varepsilon$ over a time $T$ scales as $\mathrm{polylog}(T/\varepsilon)$~\cite{Huang2026}. If the target is unphysical, such as truncated HEOM decompositions, it can still be approximated to a desired accuracy, at the cost of additional modes.

Alternative routes avoid nonconvex optimization of the Lindbladian. Reference~\cite{Huang2026} constructs coupled modes without initial guesses, from an exponential fit of the BCF followed by a convex semidefinite program (SDP) that fixes the gauge of $(h,K)$, and Ref.~\cite{Muller2026} gives an explicit construction of $(g,h,K)$ from any physical exponential BCF, together with an exponential ansatz that samples only physical BCFs. These routes avoid random initializations and can be advantageous when the number of modes grows. Like the latter ansatz, the direct parametrization used here yields a physical BCF by construction. For the two-mode fits of this work the SDP route reproduces the least-squares optimum, and we report the least-squares results.

We note that in all fits of this work, the optimized dissipation matrix is effectively of rank one: one eigenvalue of $K$ is several orders of magnitude smaller than the other. A rank-one $K$ corresponds to a single jump operator or, after a unitary change of mode basis, to a single damped mode. M\"uller and Strunz have shown that every physical BCF consisting of $N$ exponentials can be represented by $N$ coupled pseudomodes of which only one is damped, by exploiting the unitary freedom in the construction of the Lindbladian~\cite{Muller2026}. Here, this structure is not imposed: our parametrization allows a full-rank $K$, and the rank-one solution emerges from unconstrained optimization starting from random initial conditions. Because the parametrization keeps $K$ strictly positive definite, the vanishing eigenvalue can only be approached asymptotically ($x_j\to-\infty$), and the fits return a small but finite value. For the fits of this work, a single dissipative channel per bath therefore suffices, which we exploit in Sec.~\ref{sec:indep}.

\section{Quantum-trajectory circuit}
\label{sec:circuit}

Figure~\ref{fig:circuit}(b) shows the circuit that propagates one trajectory of the coupled-pseudomode dynamics, Eq.~\eqref{eq:clpm}. This section describes its components: the register and the structure of a Trotter step (Sec.~\ref{sec:circuit:register}), and the implementation of the dissipative channels (Sec.~\ref{sec:circuit:dilation}).

\subsection{Register and Trotter step}
\label{sec:circuit:register}

The register holds $\lceil\log_2\dim\mathcal{H}_{\mathrm{S}}\rceil$ qubits for the system, $n_q$ qubits per pseudomode, and a single ancilla that is reused by every dissipative channel. The Lindblad equation is unraveled into pure-state trajectories~\cite{Dalibard1992,Plenio1998}: each execution of the circuit realizes one trajectory, system observables are measured at the end, and averaging over executions reproduces the density matrix. Because each dissipative channel is implemented as a unitary followed by a measurement and reset of the ancilla, both outcomes contribute to the trajectory and no postselection is needed~\cite{Ding2024}.

The generator is split into three parts, each implemented by one type of block in Fig.~\ref{fig:circuit}(b): the coherent bath block $\mathcal{A}(\tau)$, generated by the mode Hamiltonians and the system--bath couplings of all baths; the system propagator $U_{\mathrm{el}}(\tau)=e^{-i\tau\Hel}$; and the dissipative channels $\mathcal{C}$, one per bath and jump operator, each implemented by a dilation unitary $U_{\mathrm{dil}}$ acting on the bath and the ancilla, followed by measurement and reset of the ancilla (Sec.~\ref{sec:circuit:dilation}). Each step follows the symmetric sequence $\mathcal{A}(\dt/2)\,U_{\mathrm{el}}\,\mathcal{C}\,\mathcal{A}(\dt/2)$, where consecutive half-blocks of neighboring steps fuse, so that $n_t$ steps read
\begin{equation}
  \mathcal{A}(\dt/2)\left[U_{\mathrm{el}}\,\mathcal{C}\,\mathcal{A}(\dt)\right]^{n_t-1} U_{\mathrm{el}}\,\mathcal{C}\,\mathcal{A}(\dt/2)
  \label{eq:step}
\end{equation}
and require $n_t+1$ bath blocks instead of $2n_t$. When $\mathcal{A}$ is itself a product of noncommuting sub-blocks, fusing reorders them; the difference is a splitting error of the same order as that of the step, and it vanishes when the sub-blocks commute (Sec.~\ref{sec:resources}). Since $U_{\mathrm{el}}$ and the channels act on disjoint registers, they commute exactly, and the accuracy of the step is limited by the dilated channels as discussed next.

\subsection{Dissipative channels}
\label{sec:circuit:dilation}

The dissipator of the coupled-pseudomode equation, Eq.~\eqref{eq:clpm}, contains the jump operators $L_\nu$ of Eq.~\eqref{eq:jumps}, which are linear combinations of the annihilation operators $b_j$ of the modes of one bath. Each dissipative channel $\mathcal{C}$ of the Trotter step must reproduce, over one step, the evolution generated by the dissipator $\D[L]$ of one such jump operator, which we denote $L$ in what follows. This evolution is not unitary, and can be implemented through a Stinespring dilation~\cite{Stinespring1955,nielsen2010chuang}: the jump operator, which acts on the modes of its bath, is combined with the ancilla into the Hermitian operator
\begin{equation}
  \Hdil = i\left(L\otimes\sigma - L^\dagger\otimes\sigma^\dagger\right) = A\otimes Y - B\otimes X,
  \label{eq:Hdil}
\end{equation}
where $\sigma=\ketbra{1}{0}$ acts on the ancilla, $X$ and $Y$ are Pauli operators, and $L=A+iB$ with $A$ and $B$ Hermitian. The unitary
\begin{equation}
  \Udil = e^{-i\sqrt{\dt}\,\Hdil},
  \label{eq:Udil}
\end{equation}
denoted $U_{\mathrm{dil}}^{(b)}$ for bath $b$ in Fig.~\ref{fig:circuit}(b), acts on the bath and the ancilla, which is prepared in $\ket{0}$ and then measured and reset. Tracing out the ancilla yields a channel with Kraus operators
\begin{align}
  K_0 &= \bra{0}\Udil\ket{0} = 1-\tfrac{\dt}{2}L^\dagger L + O(\dt^2),\\
  K_1 &= \bra{1}\Udil\ket{0} = \sqrt{\dt}\,L + O(\dt^{3/2}),
\end{align}
so that $\mathcal{C}\rho = K_0\rho K_0^\dagger + K_1\rho K_1^\dagger = \rho + \dt\,\D[L]\rho + O(\dt^2)$, i.e., the dilated channel reproduces the dissipative evolution over one step with an error of $O(\dt^2)$. In a trajectory, finding the ancilla in $\ket{0}$ applies the no-jump operator $K_0$, and finding it in $\ket{1}$ applies the jump operator $K_1$; both outcomes are kept, and the state is renormalized. Because $\Udil$ is unitary, the channel is exactly CPTP for any $\dt$.

The time argument of the dilation is $\sqrt{\dt}$, much larger than the step $\dt$ of the coherent blocks, which makes the dilation the numerically most demanding block. When it acts on too many qubits for exact synthesis (Sec.~\ref{sec:results}), it is implemented with a product formula of order $p$ and $r$ repetitions over the Pauli decomposition of $\Hdil$~\cite{Suzuki1991,Childs2021}. Every Pauli term of $\Hdil$ acts on the ancilla as $X$ or $Y$, so it flips the ancilla: a product of an even number of such terms is diagonal in the ancilla basis and contributes to $K_0$, and a product of an odd number contributes to $K_1$. The leading error of a product formula of order $p$ is a nested commutator of $p+1$ terms multiplied by $\tau^{p+1}$, with $\tau=\sqrt{\dt}$. For $p=1$ it has two factors and enters $K_0$ at order $\tau^2=\dt$, the same order as the dissipative term itself, so the channel is wrong at leading order and the error accumulates to $O(1)$ over a trajectory. For $p=2$ it has three factors and enters $K_1$ at order $\tau^3=\dt^{3/2}$; since $K_1$ is itself $O(\sqrt{\dt})$, it modifies the channel only at order $\dt^2$, the same order as the intrinsic error of the dilation. Second-order splitting is therefore the lowest order that preserves the accuracy of the dilated channel, and its error decreases further with $r$. These errors are quantified in Sec.~\ref{sec:errors}.

\paragraph*{Bilinear dilation.}
Because each jump operator is linear in the modes, $L=\sum_j L_j$ with $L_j=u_j b_j$, its dilation splits into one piece per mode, $\Hdil=\sum_j H_j$ with $H_j=i(L_j\otimes\sigma-L_j^\dagger\otimes\sigma^\dagger)$, each bilinear in the ladder operators of one mode and of the ancilla. The pieces share the ancilla and do not commute, $[H_i,H_j]=(L_i^\dagger L_j-L_iL_j^\dagger)\otimes Z$. A symmetric product of the pieces, each acting on a single mode and the ancilla, implements $\Udil$ with a single second-order step whose leading error enters through the ancilla in the same way as for the monolithic dilation. This variant reduces the width of the synthesized blocks and the circuit depth; on hybrid qubit--oscillator hardware, each piece is a red-sideband (Jaynes--Cummings) interaction. We use it as an alternative to the Pauli-based product formula and compare the cost and accuracy of both implementations in Sec.~\ref{sec:resources}.

\section{Structure-exploiting encoding and compilation}
\label{sec:compilation}

\subsection{Qubit encoding of the modes}
\label{sec:compilation:encoding}

Each truncated mode, $b=\sum_{n=1}^{d-1}\sqrt{n}\,\ketbra{n-1}{n}$, is encoded in $n_q=\log_2 d$ qubits by writing the Fock index in binary~\cite{Macridin2018,Sawaya2020}, and all operators are expanded in Pauli strings. The ladder operator then contains $d\log_2 d$ Pauli strings, the number operator $\nhat$ is a sum of $n_q$ single-qubit $Z$ operators, and a bilinear term $b_i^\dagger b_j$ between two modes contains up to $(d\log_2 d)^2$ strings, which makes inter-mode couplings in the Hamiltonian the dominant cost of a direct encoding (Sec.~\ref{sec:compilation:nm}). As an alternative we use the binary-reflected Gray code, $P\ket{n}=\ket{n\oplus\lfloor n/2\rfloor}$, in which adjacent Fock states differ by a single bit~\cite{Sawaya2020,DiMatteo2021}. The number of Pauli strings is unchanged, but each transition of the ladder operators flips a single qubit, which is designed to reduce the number of entangling gates. The number operator remains diagonal, now with multi-qubit $Z$ strings instead of single-qubit ones. The Gray code thus targets the \CX{} count rather than the number of non-Clifford rotations.

\subsection{Normal-mode basis}
\label{sec:compilation:nm}

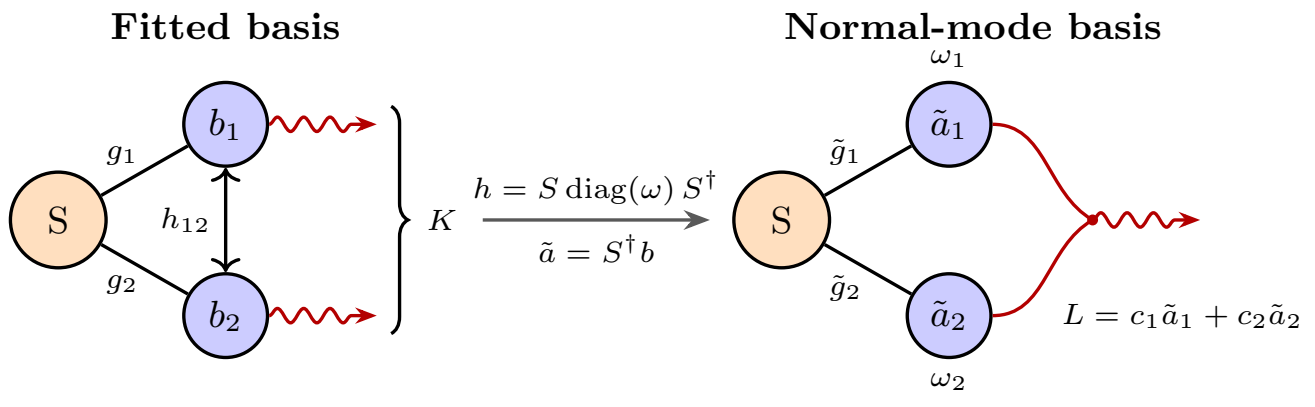
\begin{figure}[h]
\centering
\resizebox{\columnwidth}{!}{%
\begin{tikzpicture}[font=\small,
stage/.style={font=\small\bfseries, anchor=south},
  flow/.style={-{Stealth[length=2.2mm]}, thick, gray!70!black},
  sysnode/.style={circle, draw, thick, fill=orange!25, minimum size=8mm, inner sep=0pt},
  osc/.style={circle, draw, fill=blue!10, minimum size=3.2mm, inner sep=0pt},
  pm/.style={circle, draw, thick, fill=blue!20, minimum size=7mm, inner sep=0pt},
  coup/.style={decorate, decoration={snake, amplitude=0.6mm, segment length=2.2mm}},
  damp/.style={decorate, decoration={snake, amplitude=0.6mm, segment length=2.2mm, post length=1.5mm}, -{Stealth[length=1.8mm]}, thick, red!70!black},
]
\node[stage] at (1.4,2.6) {Fitted basis};
\node[sysnode] (S) at (0,1.2) {S};
\node[pm] (b1) at (1.4,2.0) {$b_1$};
\node[pm] (b2) at (1.4,0.4) {$b_2$};
\draw[thick] (S) -- node[above left, font=\scriptsize, inner sep=1pt] {$g_1$} (b1);
\draw[thick] (S) -- node[below left, font=\scriptsize, inner sep=1pt] {$g_2$} (b2);
\draw[thick, <->] (b1) -- node[left, font=\scriptsize] {$h_{12}$} (b2);
\draw[damp] (b1.east) -- ++(0.9,0) coordinate (e1);
\draw[damp] (b2.east) -- ++(0.9,0) coordinate (e2);
\draw[decorate, decoration={brace, amplitude=4pt}, thick] ($(e1)+(0.12,0.15)$) -- ($(e2)+(0.12,-0.15)$)
  node[midway, right=5pt, font=\scriptsize] {$K$};
\draw[flow] (3.55,1.2) -- node[above, font=\scriptsize, text=black, align=center] {$h=S\,\mathrm{diag}(\omega)\,S^\dagger$} node[below, font=\scriptsize, text=black] {$\tilde a=S^\dagger b$} (5.45,1.2);
\begin{scope}[shift={(6.05,0)}]
  \node[stage] at (1.6,2.6) {Normal-mode basis};
  \node[sysnode] (Sn) at (0,1.2) {S};
  \node[pm] (a1) at (1.4,2.0) {$\tilde a_1$};
  \node[pm] (a2) at (1.4,0.4) {$\tilde a_2$};
  \node[font=\scriptsize, above=0pt of a1] {$\omega_1$};
  \node[font=\scriptsize, below=0pt of a2] {$\omega_2$};
  \draw[thick] (Sn) -- node[above left, font=\scriptsize, inner sep=1pt] {$\tilde g_1$} (a1);
  \draw[thick] (Sn) -- node[below left, font=\scriptsize, inner sep=1pt] {$\tilde g_2$} (a2);
  \coordinate (J) at (2.6,1.2);
  \draw[thick, red!70!black] (a1.east) to[out=0, in=150] (J);
  \draw[thick, red!70!black] (a2.east) to[out=0, in=210] (J);
  \fill[red!70!black] (J) circle (1.3pt);
  \draw[damp] (J) -- ++(0.9,0) coordinate (Je);
  \node[font=\scriptsize] at ($(J)+(0.75,-0.8)$) {$L=c_1\tilde a_1+c_2\tilde a_2$};
\end{scope}
\end{tikzpicture}}
\caption{Normal-mode transformation scheme. In the basis of the fit, the two pseudomodes of a bath are coupled coherently through $h_{12}$ and dissipatively through the off-diagonal elements of $K$. In the eigenbasis of $h$, the coherent coupling vanishes and the modes evolve independently with frequencies $\omega_\mu$; the inter-mode coupling is carried entirely by the dissipator, whose jump operators $L_\nu=\sum_\mu c_{\mu\nu}\tilde a_\mu$ act on both modes.}
\label{fig:normalmodes}
\end{figure}

The bilinear term $\sum_{ij}h_{ij}\,b_i^\dagger b_j$ dominates the cost of a direct encoding, since each inter-mode term contains far more Pauli strings than any other term of the bath block. We therefore propose to encode the eigenmodes of $h$ rather than the original modes.
Writing
\begin{equation}
  h=S\,\mathrm{diag}(\omega)\,S^\dagger,\qquad \tilde a = S^\dagger b,
  \label{eq:nm}
\end{equation}
the mode Hamiltonian becomes $\sum_\mu\omega_\mu\nhat_\mu$, which is diagonal and compiles exactly into commuting $Z$-type rotations: single-qubit $R_z$ rotations in the binary encoding, and multi-qubit $Z$ rotations in the Gray code. The coupling vector and the jump operators transform as
\begin{equation}
  \tilde g = S^\dagger g, \qquad
  L_\nu = \sum_\mu c_{\mu\nu}\,\tilde a_\mu, \quad c_{\mu\nu} = \sqrt{\kappa_\nu}\sum_j U^{*}_{j\nu}S_{j\mu},
  \label{eq:nm_g_c}
\end{equation}
which follow from substituting $b_j=\sum_\mu S_{j\mu}\tilde a_\mu$ into the coupling term and into Eq.~\eqref{eq:jumps}. The unitary $S$ never appears as a gate, since the initial vacuum is invariant under the mode transformation and only system observables are measured. This transformation is the most effective step of the compilation pipeline as shown in Sec.~\ref{sec:resources}.

In the normal-mode basis, the coupling of each bath to the system splits into mutually commuting blocks, one per mode, each acting on the system register and a single mode. The size of each block is therefore independent of the number of modes. Each block is synthesized exactly as long as its qubits stay within the exact-synthesis threshold of Sec.~\ref{sec:results}, and with a product formula otherwise. For site-local couplings in the momentum basis, each block factorizes further into diagonal phase layers and a single-mode unitary (Sec.~\ref{sec:compilation:factorization}). The inter-mode coupling does not disappear: it moves into the dissipator, whose matrix in the normal-mode basis, $S^\dagger K S$, is in general not diagonal, so each jump operator acts on all modes of its bath (Fig.~\ref{fig:normalmodes}). This trade is favorable because the bilinear term multiplies the Pauli strings of two modes, whereas the jump operators are linear in the modes and only add them, giving at most $m\,d\log_2 d$ strings per jump. The alternative choice, diagonalizing $K$ instead of $h$, yields a single damped mode when $K$ has rank one~\cite{Muller2026}, but leaves the bilinear term in the Hamiltonian, whose Pauli strings multiply across modes; since this term enters the coherent block of every step, while the jump operators only add strings, diagonalizing $h$ is considerably cheaper.

\subsection{Momentum basis for identical baths}
\label{sec:compilation:momentum}

When the system consists of $N$ sites, each coupled through its projector $V_\alpha=\ketbra{\alpha}{\alpha}$ to its own bath, and all baths are identical and independent, we apply a discrete Fourier transform over the bath index $\alpha$, separately for each mode $j$:
\begin{equation}
  \tilde b_{kj} = \frac{1}{\sqrt N}\sum_{\alpha=0}^{N-1} e^{-2\pi i k\alpha/N}\, b_{\alpha j}, \qquad k=0,\dots,N-1.
  \label{eq:dft}
\end{equation}
Since the transformation is unitary in $\alpha$, the mode Hamiltonian $\sum_\alpha\sum_{ij}h_{ij}b_{\alpha i}^\dagger b_{\alpha j}$ and the total dissipator $\sum_{\alpha\nu}\D[L_{\alpha\nu}]$ keep their form in terms of $\tilde b_{kj}$, so the baths remain independent in $k$. The coupling becomes
\begin{equation}
  \sum_\alpha V_\alpha\, b_{\alpha j} = \sum_k \tilde V_k\, \tilde b_{kj}, \quad
  \tilde V_k = \frac{1}{\sqrt N}\sum_{\alpha} e^{\frac{2\pi i k\alpha}{N}}\ketbra{\alpha}{\alpha},
  \label{eq:Vk}
\end{equation}
and its Hermitian conjugate involves $\tilde V_k^\dagger$. The operators $\tilde V_k$ are diagonal, unitary up to the factor $N^{-1/2}$, and non-Hermitian for $k\neq0,N/2$. For $k=0$, $\tilde V_0=I/\sqrt N$: the symmetric bath couples to the identity, its dynamics factorizes from that of the system and the other baths, and it can be removed exactly. Only $N-1$ baths remain, which saves $m\,n_q$ qubits. Physically, the system dynamics is insensitive to the sum of the bath fluctuations and responds only to their differences.

The transformation is exact in the infinite Fock space, but truncation does not commute with it, since a Fock cutoff in the site basis is not a Fock cutoff in the momentum basis. The two truncated models therefore differ by an amount that vanishes as $d$ increases.

The construction requires identical baths but not uncorrelated ones. For spatially correlated baths with cyclic translation-invariant correlations of the site index, $C_{\alpha\beta}(t)=\xi(\alpha-\beta)\,C(t)$ with $\xi(0)=1$, the $k$ components remain independent with weights $w_k=\sum_\delta e^{-2\pi i k\delta/N}\xi(\delta)$, which enter as $g_k=\sqrt{w_k}\,g$. Since $\sum_k w_k=N$, the local bath is unchanged, and the $k=0$ bath still decouples for any $w_0$. All results in this paper use uncorrelated baths ($w_k=1$). For systems whose couplings are not projectors, such as spin chains with local $\sigma_z^\alpha$ couplings, $\sum_\alpha\sigma_z^\alpha$ is the total magnetization rather than the identity, and the $k=0$ bath decouples only within a fixed magnetization sector.

\begin{figure}[t]
\centering
\resizebox{\columnwidth}{!}{%
\begin{tikzpicture}[font=\small,
  stage/.style={font=\small\bfseries, anchor=south},
  flow/.style={-{Stealth[length=2.2mm]}, thick, gray!70!black},
  site/.style={circle, draw, thick, fill=orange!25, minimum size=8mm, inner sep=0pt},
  bath/.style={rounded corners=2pt, draw, thick, fill=blue!20, minimum width=20mm, minimum height=6mm, inner sep=1pt, font=\scriptsize},
  ghost/.style={rounded corners=2pt, draw, dashed, gray, fill=gray!8, minimum width=20mm, minimum height=6mm, inner sep=1pt, font=\scriptsize, text=gray},
  vdot/.style={fill=black, circle, inner sep=0pt, minimum size=2.2pt},
]
\node[stage] at (1.05,3.75) {Site basis};
\node[site] (s0) at (0,3.3) {$0$};
\node[site] (s1) at (0,2.3) {$1$};
\node[site] (sN) at (0,0.3) {\scriptsize $N\!-\!1$};
\draw[thick] (s0) -- node[left, font=\scriptsize] {$J$} (s1);
\draw[thick, dashed] (s1) -- (sN);
\node[bath] (B0) at (2.15,3.3) {bath $\alpha\!=\!0$};
\node[bath] (B1) at (2.15,2.3) {bath $\alpha\!=\!1$};
\node[bath] (BN) at (2.15,0.3) {bath $\alpha\!=\!N\!-\!1$};
\draw[thick] (s0) -- node[above, font=\scriptsize] {$V_0$} (B0);
\draw[thick] (s1) -- node[above, font=\scriptsize] {$V_1$} (B1);
\draw[thick] (sN) -- node[above, font=\scriptsize] {$V_{N-1}$} (BN);
\foreach \y in {1.55,1.3,1.05} \node[vdot] at (2.15,\y) {};
\draw[flow] (3.4,1.8) -- node[above, font=\scriptsize, text=black] {DFT} node[below, font=\scriptsize, text=black] {over $\alpha$} (4.4,1.8);
\begin{scope}[shift={(4.95,0)}]
  \node[stage] at (1.55,3.75) {Momentum basis};
  \node[draw, thick, rounded corners=3pt, fill=orange!25, minimum width=8mm, minimum height=37mm] (S) at (0,1.8) {S};
  \node[ghost] (K0) at (2.45,3.3) {bath $k\!=\!0$};
  \node[bath] (K1) at (2.45,2.3) {bath $k\!=\!1$};
  \node[bath] (KN) at (2.45,0.3) {bath $k\!=\!N\!-\!1$};
  \draw[dashed, gray] (S.east |- K0) -- node[above, font=\scriptsize, text=gray] {$\tilde V_0\propto I$} (K0);
  \draw[thick] (S.east |- K1) -- node[above, font=\scriptsize] {$\tilde V_1$} (K1);
  \draw[thick] (S.east |- KN) -- node[above, font=\scriptsize] {$\tilde V_{N-1}$} (KN);
  \foreach \y in {1.55,1.3,1.05} \node[vdot] at (2.45,\y) {};
  \node[font=\tiny, text=gray, anchor=north] at ($(K0.south)+(0,-0.02)$) {decouples};
\end{scope}
\end{tikzpicture}}
\caption{Momentum basis for $N$ sites with identical, independent baths. In the site basis, each site $\alpha$ couples through its projector $V_\alpha=\ketbra{\alpha}{\alpha}$ to its own bath. A discrete Fourier transform over the bath index, Eq.~\eqref{eq:dft}, yields $N$ independent baths labeled by $k$, which couple to the system through the diagonal phase operators $\tilde V_k$ of Eq.~\eqref{eq:Vk}. The symmetric bath $k=0$ couples through $\tilde V_0=I/\sqrt N$, decouples from the system dynamics, and is removed, leaving $N-1$ baths.}
\label{fig:momentum}
\end{figure}
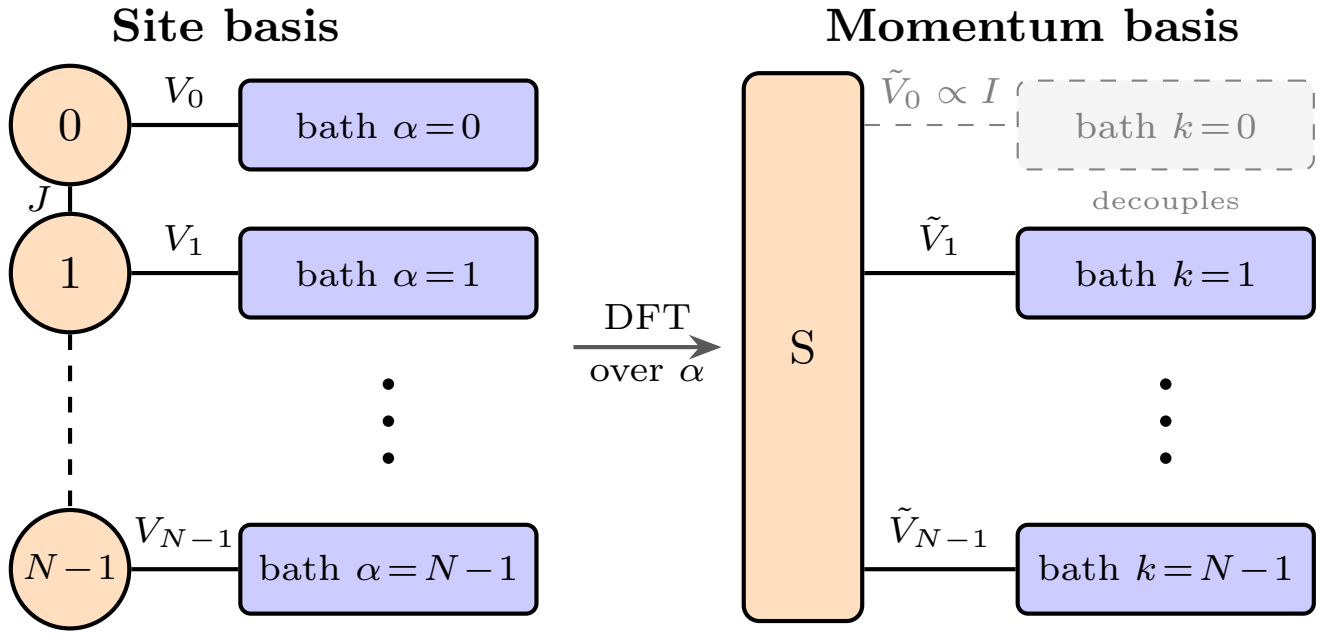

This construction also shows that $N$ identical local baths act on the system as $N-1$ independent baths, each coupled to a collective system operator. Since $\tilde V_{N-k}=\tilde V_k^\dagger$, real combinations of the baths $k$ and $N-k$ couple through the Hermitian operators $\sqrt{2/N}\sum_\alpha\cos(2\pi k\alpha/N)\ketbra{\alpha}{\alpha}$ and $\sqrt{2/N}\sum_\alpha\sin(2\pi k\alpha/N)\ketbra{\alpha}{\alpha}$, while for even $N$ the bath $k=N/2$ already couples through a Hermitian operator. The non-Hermiticity of $\tilde V_k$ is therefore a choice of representation; we use the complex form of Eq.~\eqref{eq:Vk} because all entries of $\tilde V_k$ have equal modulus, which enables the exact factorization of Sec.~\ref{sec:compilation:factorization}. For $N=2$ a single collective bath remains, coupled through $\tilde V_1=\sigma_z/\sqrt2=(\sigma_z/2)\sqrt2$: two independent baths with coupling vector $g$ are equivalent to one shared bath coupled through $\sigma_z/2$ with coupling vector $\sqrt2\,g$, i.e., with twice the bath correlation function, and with the same $h$, $K$, and $\Hel$.

\subsection{Exact factorization of the momentum-space coupling}
\label{sec:compilation:factorization}

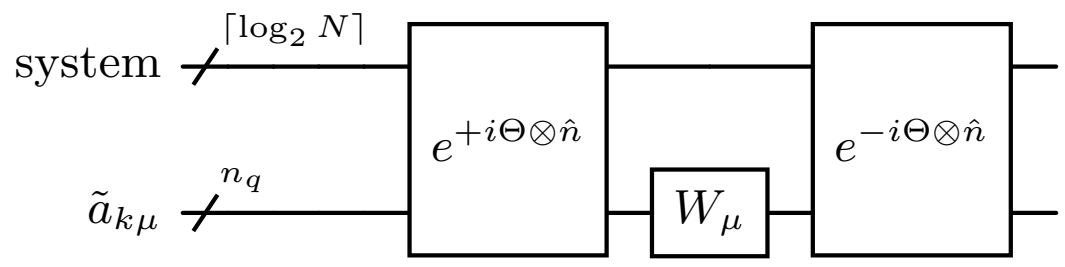
\begin{figure}[t]
\centering
\begin{quantikz}[column sep=0.3cm, row sep=0.4cm]
\lstick{system} & \qwbundle{\lceil\log_2 N\rceil} & & & & \gate[2]{e^{+i\Theta\otimes\hat n}} & & \gate[2]{e^{-i\Theta\otimes\hat n}} & \\
\lstick{$\tilde a_{k\mu}$} & \qwbundle{n_q} & & & & & \gate{W_\mu} & &
\end{quantikz}
\caption{Exact factorization of the coupling block between momentum bath $k$ and normal mode $\mu$, Eq.~\eqref{eq:factorization}. The phase layers are diagonal products of commuting $Z$-type rotations, and $W_\mu$ acts on the mode alone.}
\label{fig:factorization}
\end{figure}

After the normal-mode transformation, the coupling of bath $k$ to normal mode $\mu$ is
\begin{equation}
  H_C^{(k\mu)} = \tilde g_\mu^{*}\,\tilde V_k\,\tilde a_{k\mu} + \tilde g_\mu\,\tilde V_k^\dagger\,\tilde a_{k\mu}^\dagger.
  \label{eq:HCk}
\end{equation}
For projector-coupled sites, every $\tilde V_k$ of Eq.~\eqref{eq:Vk} is diagonal in the computational basis of the system register and multiplies each basis state $\ket{\alpha}$ by a pure phase, $\tilde V_k = N^{-1/2}e^{i\Theta}$ with $\Theta=\mathrm{diag}(2\pi k\alpha/N)$. Diagonality makes blocks with different $(k,\mu)$ commute, so each block can be implemented separately, and the pure-phase form allows each block to be factorized exactly. A phase multiplying a ladder operator can be generated by conjugation with the number operator, $e^{-i\theta\nhat}\,a\,e^{i\theta\nhat}=e^{i\theta}a$, an identity that also holds for truncated ladder operators. Applying it on each system state $\ket{\alpha}$ gives $e^{i\Theta}\otimes a = e^{-i\Theta\otimes\nhat}(I\otimes a)\,e^{i\Theta\otimes\nhat}$, and analogously for $e^{-i\Theta}\otimes a^\dagger$, so that
\begin{align}
  e^{-i\tau H_C^{(k\mu)}} &= e^{-i\Theta\otimes\nhat_{k\mu}}\left(I\otimes W_\mu\right)e^{+i\Theta\otimes\nhat_{k\mu}},
  \label{eq:factorization}\\
  W_\mu &= \exp\!\left[-\frac{i\tau}{\sqrt N}\left(\tilde g_\mu^{*}\tilde a_{k\mu} + \tilde g_\mu\tilde a_{k\mu}^\dagger\right)\right].
  \nonumber
\end{align}

The single-mode unitary $W_\mu$ acts only on the $n_q$ qubits of one mode, does not depend on the system state, and, for uncorrelated baths, is the same for every $k$; it is therefore synthesized exactly (Sec.~\ref{sec:resources}). The phase layers $e^{\pm i\Theta\otimes\nhat}$ are diagonal, since $\Theta$ is linear in the bits of the binary label $\alpha$ and $\nhat$ is a sum of $Z$-type strings. Each layer is therefore a product of commuting $Z$-type rotations, single- and two-qubit in the binary mode encoding and multi-qubit in the Gray code. Unlike $W_\mu$, the layers do not rely on generic unitary synthesis: each rotation compiles directly into \CX{} gates and one $R_z$ rotation, so that the layers are implemented without approximation for any $N$ and $d$ (Fig.~\ref{fig:factorization}).

Because the Fourier phases are linear in the bits of $\alpha$, the number of rotations in each phase layer grows only linearly with the $\lceil\log_2 N\rceil$ qubits of the system register, and the cost of each block grows as $\log N$, which results in a cost per bath that is nearly independent of $N$ (Sec.~\ref{sec:resources}). A general diagonal phase would instead require up to $2^{n_{\mathrm{S}}}-1$ rotations, and generic synthesis of the unfactorized block grows exponentially with its width. When $N$ is not a power of two, the unused basis states of the system register are never populated, since all terms of the Hamiltonian preserve the physical subspace.

\section{Results}
\label{sec:results}

We apply the method to two classes of models. Section~\ref{sec:shared} validates the coupled-pseudomode circuit on a two-level system coupled to a single bath, for overdamped and underdamped spectral densities. Section~\ref{sec:indep} turns to chains of sites with independent baths, where the transformations of Sec.~\ref{sec:compilation} apply; there we validate the optimized circuit at $N=2$ and $N=3$, analyze its error budget, and determine the fault-tolerant resources and their scaling with $N$. We first summarize the conventions and numerical methods common to all calculations.

Energies are given in units of $E_u=500$~cm$^{-1}$ and times in units of $\hbar/E_u\approx10.6$~fs. The system Hamiltonians, spectral densities, pseudomode fits, Fock truncations, time steps, and simulated time windows are given separately for each model. The pseudomodes always start in the vacuum, and system populations are averaged over $N_{\mathrm{tr}}$ trajectories, with standard error $\sigma/\sqrt{N_{\mathrm{tr}}}$.

Blocks acting on at most five qubits are compiled by exact unitary synthesis~\cite{Shende2006}, whose generic cost is about $\frac{23}{48}4^w$ \CX{} gates for a $w$-qubit block, structured blocks being cheaper. Wider blocks are implemented with a second-order Suzuki formula~\cite{Suzuki1991} with $r$ repetitions. All circuits are transpiled with Qiskit~2.2.3~\cite{Qiskit2024} at optimization level 3.

The circuits are emulated classically in two modes. In \emph{deterministic} mode, the density matrix of the register is propagated through the same blocks as the circuit, and each dissipative channel is applied as the exact average over both outcomes of its dilated ancilla, $\rho\to K_0\rho K_0^\dagger+K_1\rho K_1^\dagger$, so that the approximations of the circuit are isolated from sampling noise. Replacing these Kraus maps by a fourth-order Runge--Kutta integration of the dissipator, while keeping all unitary blocks and the step structure unchanged, yields the reference used in the error analysis of Sec.~\ref{sec:errors}. In \emph{trajectory} mode, the state vector is propagated through the circuit, and each ancilla measurement outcome is drawn at random with its quantum-mechanical probability, as on hardware. Because the density matrix grows as $4^{n}$ with the number of qubits $n$, against $2^{n}$ for the state vector, the deterministic mode is restricted to small registers.

The emulated trajectories are statistically equivalent to hardware shots, since the ancilla outcomes are drawn from the same distribution. The emulation, however, has access to the full state vector, so that the system populations of each trajectory are recorded as expectation values and at all times of a single trajectory. On hardware, each shot instead yields a single binary outcome from a final measurement of the system register, and each point in time requires its own set of shots.

In both modes, each block is applied as a dense matrix built from its circuit; blocks implemented with a product formula are expanded into that formula before the matrix is built, so that the emulation reproduces the circuit rather than the exact exponential. The gate counts of Sec.~\ref{sec:resources} are obtained from the transpiled circuits of the same blocks, and applying these transpiled circuits instead of the dense matrices reproduces the same dynamics. The dense matrices are therefore used for efficiency, and this agreement verifies the compiled circuits on which the resource estimates are based. Trajectories are distributed over MPI processes, each propagating independent trajectories.

\subsection{Validation on a shared bath}
\label{sec:shared}

We consider a two-level system with $\Hel=\Delta\,\sigma_x$, $\Delta=1$, initially in the state $\ket{0}$, coupled to a single bath through $V=\sigma_z/2$. The same Hamiltonian describes the spin-boson model and an excitonic dimer whose two site energies couple, with opposite signs, to a common bath coordinate. We use the overdamped Drude--Lorentz and the underdamped Brownian spectral densities,
\begin{align}
  J_{\mathrm{D}}(\omega) &= \frac{2\lambda\gamma\,\omega}{\omega^2+\gamma^2},
  \label{eq:JD}\\
  J_{\mathrm{B}}(\omega) &= \frac{2\lambda\gamma\,\omega_0^2\,\omega}{(\omega_0^2-\omega^2)^2+\gamma^2\omega^2},
  \label{eq:JB}
\end{align}
both normalized so that the reorganization energy is $\lambda=\pi^{-1}\int_0^\infty d\omega\,J(\omega)/\omega$.

The parameters are adapted from the perylene bisimide (PBI) dimer of Ref.~\cite{Cainelli2022}, which we study with independent baths in Sec.~\ref{sec:indep}: $\omega_0=2.8$ ($1400$~cm$^{-1}$) and $\gamma=1$ ($500$~cm$^{-1}$), at $T=300$~K. For the validation, we reduce the reorganization energy from $\lambda=2.4$ to $\lambda=1$ ($500$~cm$^{-1}$) and use the same $\lambda$ and $\gamma$ for both spectral densities, so that the two cases differ only in the shape of $J(\omega)$: a broad overdamped bath and a peaked underdamped one with an oscillating BCF. By the equivalence of Sec.~\ref{sec:compilation:momentum}, a shared bath with BCF $C(t)$ coupled through $\sigma_z/2$ is equivalent to a dimer with two independent baths, each with BCF $C(t)/2$, so these models probe a weaker coupling than the independent-bath model of Sec.~\ref{sec:indep}. 
For each spectral density, the HEOM reference uses the poles of $J(\omega)$ together with two Pad\'e terms of the Bose function~\cite{Hu2010}, with the hierarchy converged at depth $\mathcal{N}_{\max}=12$, and $m=2$ coupled pseudomodes are fitted to this decomposition on 100 (Drude--Lorentz) and 50 (Brownian) uniformly spaced points in $t\in[0,10]$, following the procedure of Sec.~\ref{sec:model:fit}.

These calculations use a simpler version of the circuit, without the step structure of Sec.~\ref{sec:circuit:register} and the transformations of Sec.~\ref{sec:compilation}. The full coherent Hamiltonian $H$ of Eq.~\eqref{eq:clpm} is propagated as a single block with a second-order product formula over its Pauli decomposition, followed by the dilated channels, implemented as in Sec.~\ref{sec:circuit:dilation} with a second-order product formula with $r=2$ repetitions; the pseudomodes are encoded in the basis of the fit with a binary encoding, and all dissipative channels are retained. As a reference for the circuit, the coupled-pseudomode Lindblad equation is integrated directly with a fourth-order Runge--Kutta scheme. The difference between this reference and HEOM measures the error of the fit alone, while the difference between the circuit trajectories and the reference contains the errors of the circuit and the sampling noise; a detailed decomposition of these errors is given in Sec.~\ref{sec:errors}. Each mode is truncated at $d=8$ Fock states, and $N_{\mathrm{tr}}=1\times10^4$ trajectories are propagated up to $t=10$ ($\approx106$~fs) with time steps $\dt=2\times10^{-2}$, $1\times10^{-2}$, and $5\times10^{-3}$ for both spectral densities.

\begin{figure*}[t]
\centering
\includegraphics[width=\textwidth]{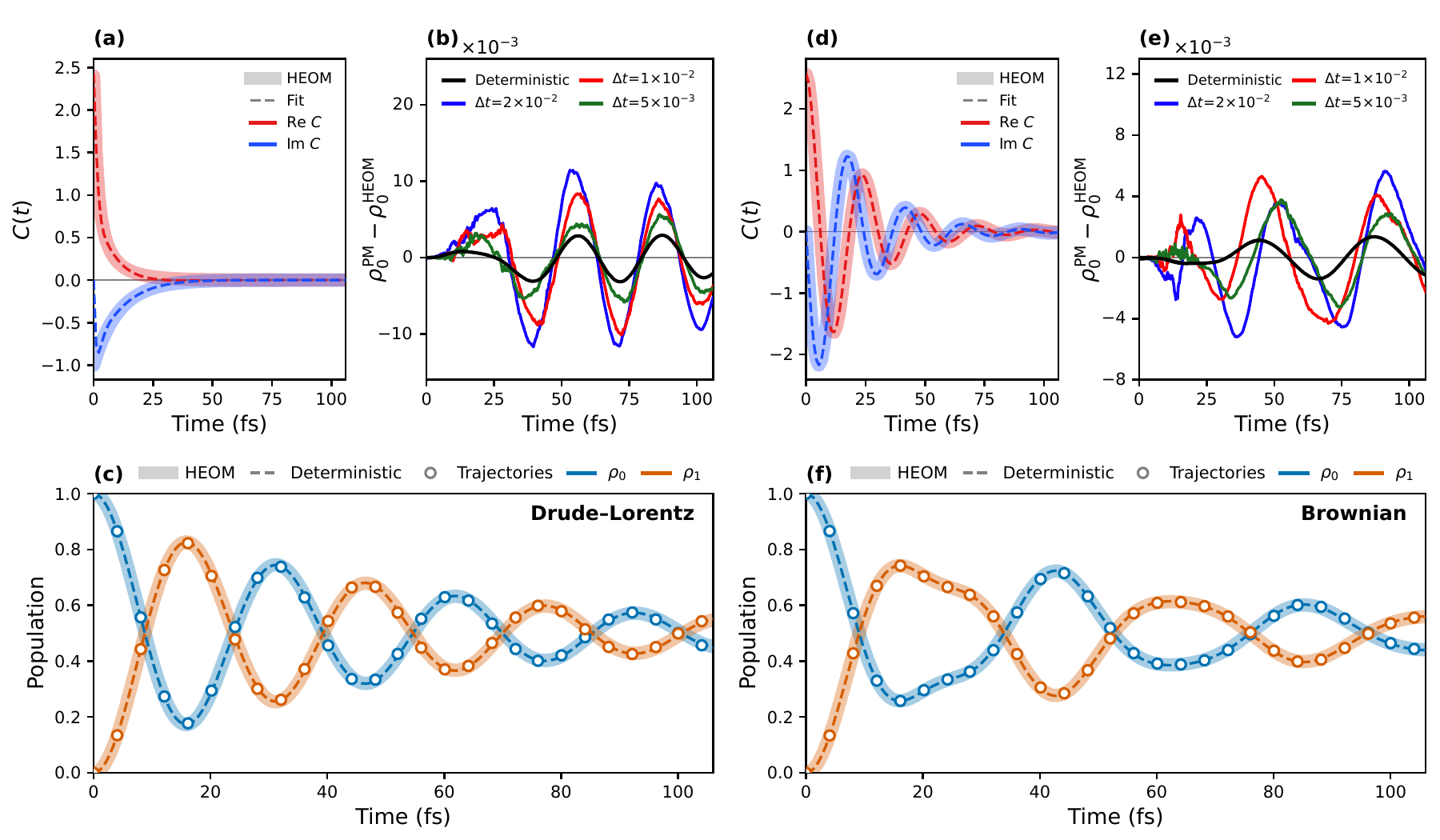}
\caption{Validation on a shared bath at $T=300$~K, for a Drude--Lorentz (left) and a Brownian (right) spectral density. (a,d) Real and imaginary parts of the BCF: HEOM decomposition and coupled-pseudomode fit. (b,e) Deviation of $\rho_0$ from HEOM for the deterministic propagation and for trajectories with several time steps, averaged over $1\times10^{4}$ trajectories. (c,f) Populations from HEOM, from the direct integration of the coupled-pseudomode Lindblad equation (deterministic), and from circuit trajectories with $\dt=2\times10^{-2}$.}
\label{fig:shared}
\end{figure*}

Figure~\ref{fig:shared} summarizes the results for the shared baths. The Drude--Lorentz BCF requires special care [Fig.~\ref{fig:shared}(a)]. Because $J_{\mathrm{D}}(\omega)$ decays only as $1/\omega$, its BCF is singular at $t=0$: $\mathrm{Im}\,C(0^{+})=-\lambda\gamma\neq0$, and $\mathrm{Re}\,C(0)$ diverges logarithmically. A finite set of pseudomodes yields a BCF that is continuous at $t=0$ (Sec.~\ref{sec:model:fit}), so the fit approximates the short-time behavior with a rapidly decaying exponential component (fast mode). Two coupled modes reproduce both parts of the BCF, except in the immediate vicinity of $t=0$, where the fitted imaginary part starts from zero and reaches the target through the fast mode within about one femtosecond. The Brownian BCF [Fig.~\ref{fig:shared}(d)] is instead continuous at $t=0$, with $\mathrm{Im}\,C(0)=0$; its dominant term oscillates with frequency $\tilde\omega=\sqrt{\omega_0^2-\gamma^2/4}$ and decays at rate $\gamma/2$, while the low-temperature corrections decay faster, and two coupled modes reproduce it without any fast mode.

The populations [Figs.~\ref{fig:shared}(c) and~\ref{fig:shared}(f)] show damped coherent oscillations between the two states. They are damped more slowly for the Brownian bath, whose spectral weight is concentrated around $\omega_0$ rather than at the low frequencies that dominate dephasing, and whose dynamics also shows a shoulder near 25~fs due to the coupling to the underdamped vibration. For both spectral densities, the direct integration of the coupled-pseudomode model and the circuit trajectories with $\dt=2\times10^{-2}$ follow the HEOM dynamics over the whole time window.

The deviation of the direct integration from HEOM, which measures the error of the fit, remains below $3\times10^{-3}$ for the Drude--Lorentz bath and $1.4\times10^{-3}$ for the Brownian bath [Figs.~\ref{fig:shared}(b) and~\ref{fig:shared}(e)]. The singular behavior of the Drude--Lorentz BCF leaves no visible error at short times and the shape of the deviation differs between the two baths. For the Drude--Lorentz bath, it vanishes at the maxima and minima of the populations and is largest where $\rho_0$ and $\rho_1$ cross, which corresponds to a slight shift in the frequency of the coherent oscillations. For the Brownian bath, it vanishes where the populations cross and is largest at their maxima and minima, positive at the maxima of $\rho_0$ and negative at its minima, which corresponds to slightly weaker damping of the oscillations.

The deviation of the trajectories is larger, up to $1.2\times10^{-2}$ for the Drude--Lorentz bath and $5.6\times10^{-3}$ for the Brownian bath at $\dt=2\times10^{-2}$. The statistical uncertainty was not recorded for these runs; it is bounded by $1/(2\sqrt{N_{\mathrm{tr}}})\approx5\times10^{-3}$, and repeating the Brownian calculation with $\dt=2\times10^{-2}$ changes the deviation by a few $10^{-3}$, which indicates the size of the sampling noise. For the Drude--Lorentz bath, the deviation decreases systematically as $\dt$ is reduced, to $6\times10^{-3}$ at $\dt=5\times10^{-3}$, approaching the error of the fit, as expected for the first-order splitting between the coherent and dissipative steps. For the Brownian bath, the circuit error is smaller and comparable to the sampling noise, so that no systematic trend with $\dt$ can be resolved with this number of trajectories. To resolve contributions of this size, the independent-bath calculations of Sec.~\ref{sec:indep} use up to $5\times10^5$ trajectories and record the statistical uncertainty of each run.

These results establish that two coupled pseudomodes per bath, implemented with a single reusable ancilla and without postselection, reproduce the HEOM dynamics for both spectral densities within the accuracy of the fit. Having validated the method, we now apply the optimized circuit to the PBI dimer and trimer with independent baths (Sec.~\ref{sec:indep}), where we analyze the structure-exploiting transformations, the sources of numerical error, and the implementation cost.

\subsection{Independent baths}
\label{sec:indep}

We consider $N$ sites forming an open chain,
\begin{equation}
  \Hel = J\sum_{\alpha=0}^{N-2}\left(\ketbra{\alpha}{\alpha+1}+\ketbra{\alpha+1}{\alpha}\right),
  \label{eq:Hel}
\end{equation}
with $J=1$ and equal site energies, so that $\Hel=J\sigma_x$ for $N=2$. Each site couples through its projector $V_\alpha=\ketbra{\alpha}{\alpha}$ to its own bath, all baths being identical and uncorrelated. This is the Frenkel-exciton model of energy transfer~\cite{Ishizaki2009}, here with the parameters of the PBI aggregates of Ref.~\cite{Cainelli2022}. Each bath has the Brownian spectral density of Eq.~\eqref{eq:JB} with $\omega_0=2.8$ ($1400$~cm$^{-1}$), $\lambda=2.4$ ($1200$~cm$^{-1}$), and $\gamma=1$ ($500$~cm$^{-1}$), at $T=300$~K. We study the dimer ($N=2$) and the trimer ($N=3$), starting with the excitation on the first site, $\ket{0}$, and the pseudomodes in the vacuum.

\begin{figure*}[t]
\centering
\includegraphics[width=\textwidth]{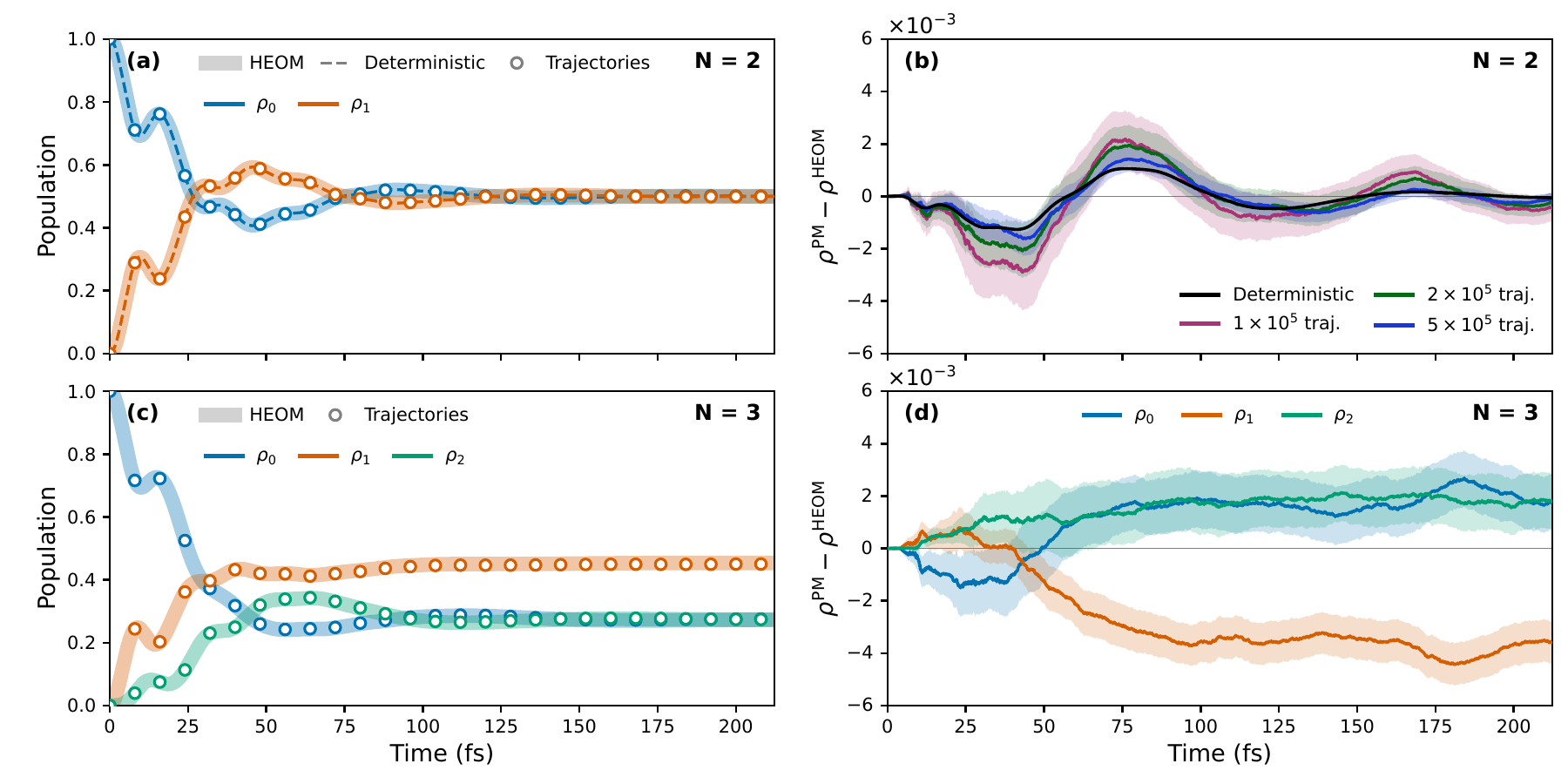}
\caption{(a,c) Site populations for $N=2$ and $N=3$, from HEOM (shaded line), from the deterministic propagation without dilation ($N=2$ only, dashed line), and from circuit trajectories (circle markers). (b) Deviation of $\rho_0$ from HEOM for $N=2$: deterministic propagation and averages over $1\times10^5$, $2\times10^5$, and $5\times10^5$ trajectories. (d) Deviation of each site population from HEOM for $N=3$, averaged over $1\times10^5$ trajectories. Shaded bands give $\pm2\sigma/\sqrt{N_{\mathrm{tr}}}$.}
\label{fig:indep}
\end{figure*}

As in Sec.~\ref{sec:shared}, the HEOM reference uses the poles of $J(\omega)$ together with two Pad\'e terms of the Bose function, and the hierarchy is converged at depth $\mathcal{N}_{\max}=12$. Since the BCF scales with the reorganization energy, we fit $m=2$ coupled modes to the decomposition for $\lambda=2.4$ with the procedure and the time grid of Sec.~\ref{sec:shared}. The fitted correlation function is a sum of two complex exponentials, one reproducing the Brownian pole and one effective low-temperature pole with a negative amplitude, which independent modes could not represent [Eq.~\eqref{eq:Clorentz}]. The fit thus compresses the four terms of the HEOM decomposition into two exponentials, represented by two coupled modes. The Fock truncation is considered to be converged at $d=8$ ($n_q=3$): the difference with respect to $d=16$ is of the order of $10^{-6}$ in the populations.

As anticipated in Sec.~\ref{sec:model:fit}, the dissipation matrix of the fit is effectively of rank one: one of its eigenvalues is several orders of magnitude smaller than the other. Dissipative channels with $\kappa_\nu<\kappa_{\mathrm{cut}}=10^{-3}$ are discarded, which removes the negligible channel associated with the small eigenvalue and leaves a single dissipative channel per bath, halving the number of dilations and mid-circuit measurements per step. In the normal-mode basis, the remaining jump operator acts on both modes of each bath.

These calculations use all the transformations of Sec.~\ref{sec:compilation}: the momentum basis, which leaves $N-1$ baths, the normal-mode basis, the exact factorization of the coupling blocks, and the Gray code, together with merged half-blocks and the channel cutoff. The coherent blocks are synthesized exactly, and the dilation is implemented with a second-order product formula with $r=2$ repetitions. The dynamics is propagated up to $t=20$ ($\approx212$~fs) with $\dt=1\times10^{-3}$ ($n_t=20\,000$ steps). The deterministic mode, whose cost grows with the square of the register dimension, is used only for $N=2$ (8 qubits), where it serves as a reference for trajectory averages over $N_{\mathrm{tr}}=1\times10^5$, $2\times10^5$, and $5\times10^5$ and for the error analysis of Sec.~\ref{sec:errors}. For $N=3$ (15 qubits), the deterministic propagation becomes impractical, and only trajectories are used, with $N_{\mathrm{tr}}=1\times10^5$.

\subsubsection{Population dynamics}
\label{sec:indep:results}

For each time we record the mean population over $N_{\mathrm{tr}}$ trajectories, its standard deviation $\sigma$, and the standard error $\sigma/\sqrt{N_{\mathrm{tr}}}$. Each MPI process draws from its own random stream, with seeds chosen such that no stream is shared between runs; configurations that are compared directly use the same seeds, so that their difference is not masked by sampling noise. The sampling noise is correlated in time, so that a single run deviates smoothly from the mean and can resemble a small phase or amplitude shift, which should not be mistaken for a systematic error.

\begin{figure*}[t]
\centering
\includegraphics[width=\textwidth]{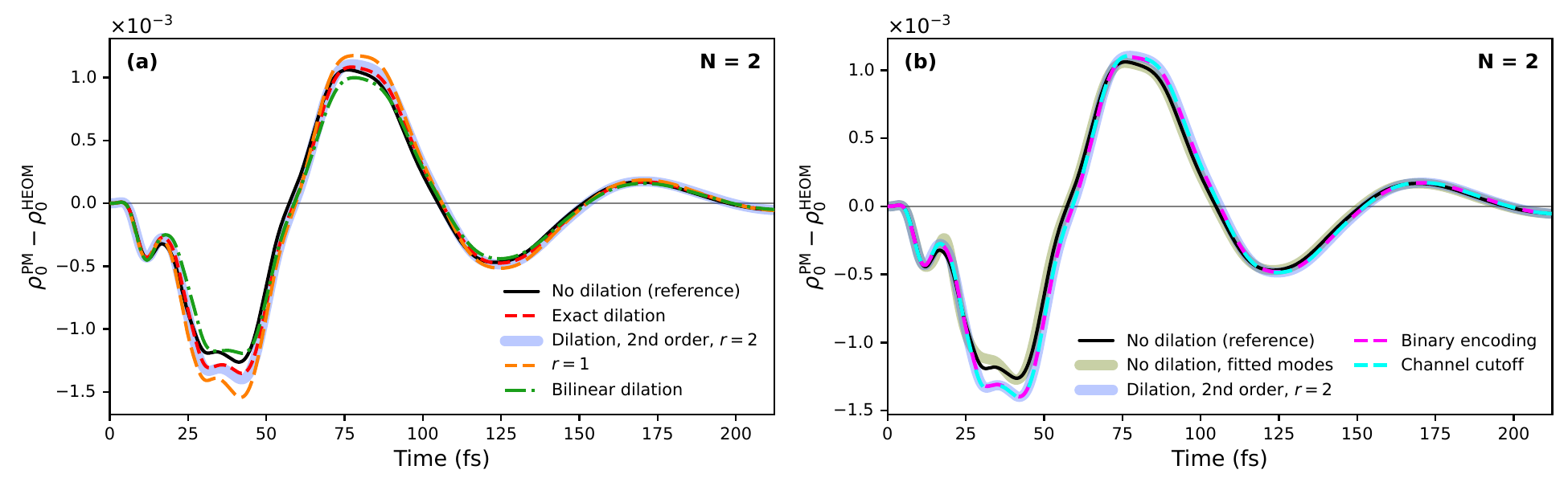}
\caption{Error budget for $N=2$: deviation of $\rho_0$ from HEOM for deterministic propagations with $\dt=1\times10^{-3}$. (a) Implementation of the dilation: reference without dilation, exact dilation, and dilations implemented with a second-order product formula with $r=2$ and $r=1$ repetitions or split into per-mode pieces (bilinear dilation). (b) Transformations and simplifications: reference without dilation with and without the normal-mode transformation (fitted modes), and the $r=2$ configuration with binary instead of Gray encoding or with the channel cutoff.}
\label{fig:errors}
\end{figure*}

Figure~\ref{fig:indep} shows the populations for $N=2$ and $N=3$ [(a) and (c)] and their deviation from HEOM [(b) and (d)], where shaded bands give $\pm2\sigma/\sqrt{N_{\mathrm{tr}}}$. For $N=2$ [Fig.~\ref{fig:indep}(a)], the population of the initially excited site is modulated during the first 25~fs by the underdamped vibration, as in the shared-bath case, and then relaxes through damped oscillations to equal populations, as required by the symmetry of the dimer. The deterministic propagation and the trajectories follow the HEOM dynamics over the whole time window. The deviation of the deterministic propagation from HEOM [Fig.~\ref{fig:indep}(b)] stays below $1.3\times10^{-3}$. The trajectory averages converge to it as the number of trajectories increases, from a maximum deviation of $2.9\times10^{-3}$ with $1\times10^5$ trajectories to $1.6\times10^{-3}$ with $5\times10^5$, and the deterministic result remains within about their statistical uncertainty. The residual difference is consistent with the sampling noise, which is correlated in time; the errors of the dilation and of the other circuit approximations, which the deterministic reference does not include, are analyzed in Sec.~\ref{sec:errors}. At long times the deviation vanishes, since the populations relax to $1/2$ independently of the bath.

For $N=3$ [Fig.~\ref{fig:indep}(c)], where the deterministic propagation is no longer feasible, the trajectories reproduce the HEOM dynamics, including the relaxation toward a state in which the central site is more populated, consistent with the thermalization of the chain, whose ground state has the largest weight on the central site. Unlike for $N=2$, the long-time populations now depend on the system--bath equilibrium. After about 50~fs, the trajectories develop a deviation of about $-3.6\times10^{-3}$ for the central site, compensated by $+1.8\times10^{-3}$ on each outer site [Fig.~\ref{fig:indep}(d)], as required by the conservation of the total population and the symmetry of the chain. This deviation persists over time and, since the equilibrium populations are sensitive to the representation of the bath, we attribute this deviation mainly to the error of the fit, comparable to the error of the fit found for $N=2$ and for the shared bath, which no longer averages out at long times as it does for the symmetric dimer.

\subsubsection{Error Isolation}
\label{sec:errors}

Numerical errors in this problem are signed and can cancel, so comparing each approximation only with HEOM can be misleading. We therefore compare each approximation against a numerical reference with the same pseudomode model and the same step structure, in which the dilated channels are replaced by a Runge--Kutta integration of the dissipator and all unitary blocks, in the normal-mode basis, are synthesized exactly. The difference between this reference and the dilated circuit isolates the contribution of the Stinespring dilation, while the deviation of the reference from HEOM is dominated by the error of the fit, with a smaller contribution from the splitting between blocks. Each further approximation is then switched on individually, and its effect is measured against the configuration that differs only in that approximation. We report the maximum absolute population difference over the simulated time window.

\begin{figure*}[t]
\centering
\includegraphics[width=\textwidth]{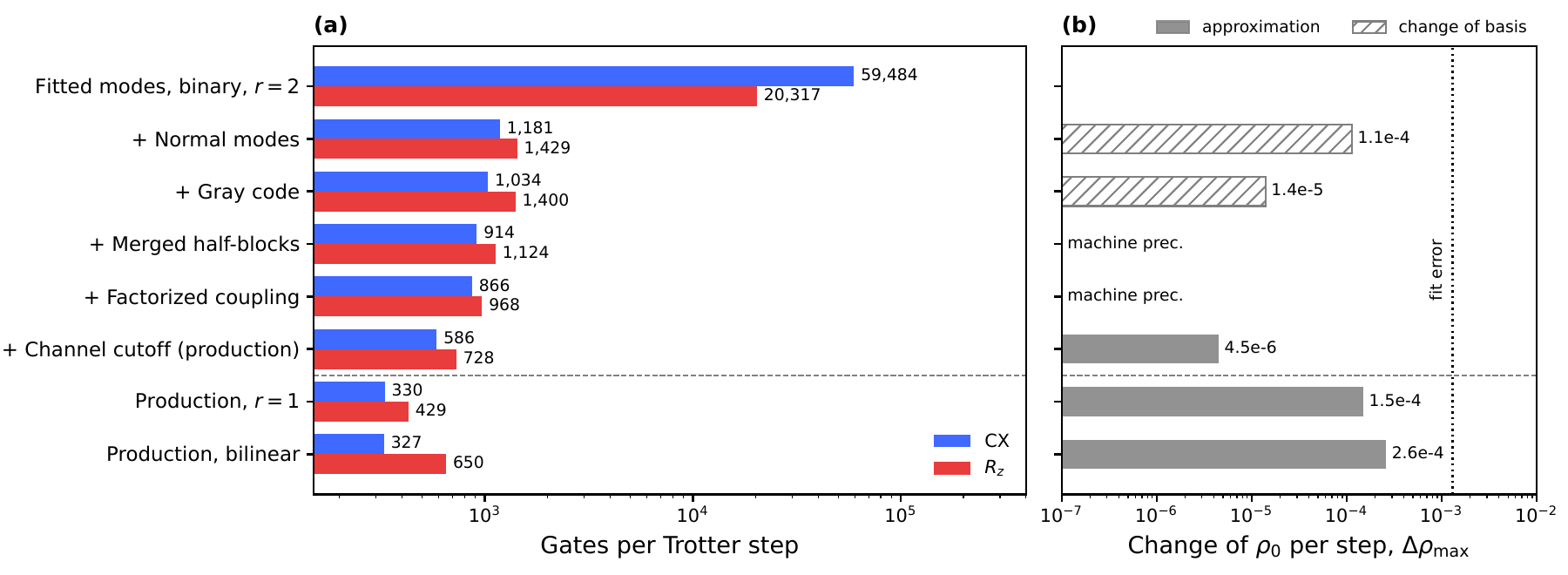}
\caption{Cost and accuracy along the compilation pipeline for $N=2$. (a) \CX{} gates and non-Clifford $R_z$ rotations per Trotter step; optimizations are cumulative down to the production configuration, and the last two rows, below the dashed line, modify only its dilation. (b) Maximum change of $\rho_0$ caused by each step alone, with respect to the configuration above or, for the last two rows, to the production configuration; the changes are signed and can partially cancel, so their sum only bounds the combined effect. Hatched bars mark exact changes of basis, whose effect arises from the different discretization of the dynamics, and the dotted line marks the error of the pseudomode fit, which dominates the deviation of all configurations from HEOM.}
\label{fig:pipeline}
\end{figure*}

Figure~\ref{fig:errors} shows the deviation from HEOM of each configuration, with the implementations of the dilation in panel (a) and the remaining transformations and simplifications in panel (b); the contribution of each approximation is given below and, together with its cost, in Sec.~\ref{sec:resources}. The largest error is that of the pseudomode fit, which, as stated in Sec.~\ref{sec:indep:results}, stays below $1.3\times10^{-3}$, while all numerical approximations are several times smaller.

The Stinespring dilation [Fig.~\ref{fig:errors}(a)] contributes less than $1.6\times10^{-4}$ when it is implemented exactly. Although not explicitly shown here, in a separate test the local error of a single dilated channel decreases by a factor of approximately $4$ when $\dt$ is halved, confirming the $O(\dt^2)$ error derived in Sec.~\ref{sec:circuit:dilation}. Implementing the dilation with a second-order product formula adds a maximum error below $5\times10^{-5}$ with $r=2$ repetitions and $2\times10^{-4}$ with $r=1$, and the bilinear dilation adds $2.1\times10^{-4}$, similar to the monolithic dilation with one repetition, since both retain the dominant commutator through the ancilla. The bilinear curve lies closer to HEOM than the $r=2$ curve in Fig.~\ref{fig:errors}(a) only because its error has the opposite sign to that of the fit and partially cancels it. A first-order splitting of the dilation, in contrast, fails within the first steps, as anticipated by the parity argument of Sec.~\ref{sec:circuit:dilation}, and is therefore not shown.

The transformations and simplifications of Sec.~\ref{sec:compilation} [Fig.~\ref{fig:errors}(b)] have smaller effects. The normal-mode transformation is an exact change of basis, but the reference without it (fitted modes) differs from the reference by $1.1\times10^{-4}$, through the different splitting of the step and the Fock truncation, which does not commute with the mode transformation. The Gray code is an exact relabeling of the Fock states, replacing it by the binary encoding changes the populations of the $r=2$ configuration by $1.4\times10^{-5}$, through the different Pauli decomposition of the Trotterized dilation, and the channel cutoff changes them by $4.5\times10^{-6}$, so that the binary-encoding and channel-cutoff curves coincide with the $r=2$ curve on the scale of the figure. The factorization of the coupling blocks is exact (Sec.~\ref{sec:compilation:factorization}), and the compiled circuits reproduce the dynamics of the dense blocks. In total, the numerical approximations of the production configuration contribute $1.9\times10^{-4}$, several times less than the error of the fit, while the statistical error of $1\times10^5$ trajectories is of order $1\times10^{-3}$, comparable to the latter.

\subsubsection{Resources and scaling}
\label{sec:resources}

Gate counts are obtained per Trotter step from the transpiled circuits, assuming all-to-all connectivity. Two-qubit counts and depth refer to the $\{u,\CX\}$ basis. Non-Clifford rotations are counted after recompiling into the Clifford+$R_z$ basis $\{R_z,\sqrt{X},X,\CX\}$ and keeping the $R_z$ gates whose angles are not multiples of $\pi/2$, the standard metric for fault-tolerant cost~\cite{wecker2014troyer,babbush2018chan,Lee2021}. Each rotation is converted into T gates with the Ross--Selinger estimate of $3\log_2(1/\epsilon)$ T gates per rotation~\cite{Ross2016}. We use $\epsilon=10^{-11}$, which costs 110 T gates per rotation and, for $N=2$, bounds the accumulated synthesis error over a trajectory of $n_t=20\,000$ steps below $2\times10^{-4}$ even if the errors of the repeated rotations add coherently. For larger systems, $\epsilon$ must decrease in proportion to the number of rotations per trajectory, which increases the T count only logarithmically. We also report the number of mid-circuit measurements per step and the total number of qubits,
\begin{equation}
  n_{\mathrm{qubits}} = \lceil\log_2\dim\mathcal{H}_{\mathrm{S}}\rceil + m\,n_q\,B + 1,
  \label{eq:nqubits}
\end{equation}
for $B$ baths of $m$ modes each, with $B=N-1$ in the momentum basis. Every compiled block acting on at most ten qubits is checked against its exact matrix.

\begin{table*}[!t]
\caption{Resources for the factorized coupling with the channel cutoff and $r=2$ ($r=1$ in parentheses): logical qubits, \CX{} gates and non-Clifford rotations per step, and T gates per trajectory of $n_t=20\,000$ steps for $\epsilon=10^{-11}$. The two-qubit depth per step is 0.86--0.98 times the \CX{} count, and each step contains $N-1$ mid-circuit measurements, one per bath.}
\label{tab:endtoend}
\begin{ruledtabular}
\begin{tabular}{rrrrr}
$N$ & Qubits & \CX{}/step & $R_z$/step & T/trajectory \\
\hline
2  & 8   & 586 (330)          & 728 (429)          & $1.6\times10^{9}$ ($9.4\times10^{8}$) \\
3  & 15  & 1\,294 (782)       & 1\,511 (913)       & $3.3\times10^{9}$ ($2.0\times10^{9}$) \\
4  & 21  & 1\,800 (1\,032)    & 2\,218 (1\,321)    & $4.9\times10^{9}$ ($2.9\times10^{9}$) \\
8  & 46  & 4\,378 (2\,586)    & 5\,281 (3\,188)    & $1.2\times10^{10}$ ($7.0\times10^{9}$) \\
16 & 95  & 9\,947 (6\,107)    & 11\,799 (7\,314)   & $2.6\times10^{10}$ ($1.6\times10^{10}$) \\
32 & 192 & 21\,789 (13\,853)  & 25\,268 (15\,999)  & $5.5\times10^{10}$ ($3.5\times10^{10}$) \\
\end{tabular}
\end{ruledtabular}
\end{table*}

\begin{figure*}[!t]
\centering
\includegraphics[width=\textwidth]{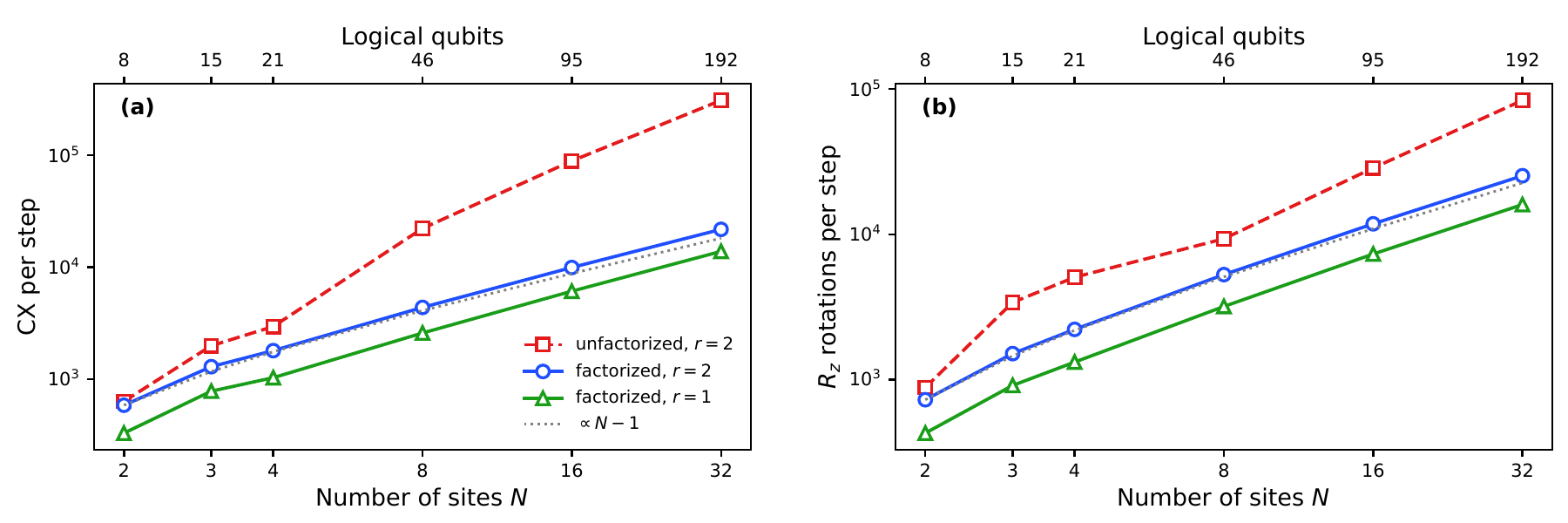}
\caption{Cost per Trotter step as a function of the number of sites $N$, in the momentum basis with the channel cutoff: (a) \CX{} gates and (b) non-Clifford $R_z$ rotations, for the unfactorized and factorized coupling with $r=2$ dilation repetitions, and for the factorized coupling with $r=1$. The top axis gives the number of logical qubits, and the dotted line is proportional to the number of baths, $N-1$, normalized to the factorized $r=2$ configuration at $N=2$.}
\label{fig:scaling}
\end{figure*}

Figure~\ref{fig:pipeline} follows the compilation pipeline for $N=2$, from a direct encoding of the fitted modes to the production configuration, and shows for each step its cost [(a)] and its effect on the populations [(b)], taken from the error budget of Sec.~\ref{sec:errors}. The normal-mode transformation is by far the most effective step: it removes the inter-mode Pauli strings of the bilinear term and reduces the cost per step from $59\,484$ to $1\,181$ \CX{} gates, a factor of about 50, and from $20\,317$ to $1\,429$ rotations. Since the direct encoding uses the same split step structure as the other configurations, it applies the expensive mode block twice per bath block; a step that applied it only once would cost about half as much, so that the reduction in \CX{} gates attributable to the normal modes alone is closer to a factor of 25. The Gray code, the merging of half-blocks, the factorization of the coupling, and the channel cutoff reduce the cost further. The normal-mode transformation and the Gray code are exact changes of basis, but they modify the discretization of the dynamics: in the normal-mode basis the Trotter step is split into different blocks and the Fock truncation, which does not commute with the mode transformation, acts on different modes, which changes the populations by $1.1\times10^{-4}$; the Gray code changes the Pauli decomposition of the Trotterized dilation, and hence its product-formula error, which changes them by $1.4\times10^{-5}$. Neither change is an approximation of the model, and both are small compared with the error of the fit. The factorization is exact, and the merged and unmerged sequences agree to machine precision, so that the channel cutoff, with an effect of $4.5\times10^{-6}$, is the only approximation among these steps. 

The production configuration requires 586 \CX{} gates and 728 rotations per step, a reduction by factors of about 100 and 28 with respect to the direct encoding with the same step structure, while all configurations deviate from HEOM by $1.2$--$1.5\times10^{-3}$, close to the error of the fit. With the channel cutoff, a single mid-circuit measurement per step suffices, and the compilation is exact: every compiled block reproduces the elements of its exact matrix to $3\times10^{-9}$. Without the cutoff, the transpiler removes the very small rotations of the negligible channel, so that the compiled blocks slightly approximate their exact matrices and the gate counts of those configurations are slightly underestimated.

The production configuration uses $r=2$ repetitions in the dilation. A single repetition reduces the cost by a further factor of about 1.7, to 330 \CX{} gates and 429 rotations per step, and changes the populations by $1.5\times10^{-4}$, several times less than the error of the fit and the sampling noise of $1\times10^5$ trajectories; when these dominate, $r=1$ is an equally accurate and cheaper choice. The bilinear dilation, with an accuracy comparable to that of $r=1$, requires the same number of \CX{} gates and has the smallest two-qubit depth (246 per step, against 318 for $r=1$), but needs more rotations (650). It is therefore preferable where circuit depth is the limiting resource, whereas $r=1$ minimizes the non-Clifford cost.

Figure~\ref{fig:scaling} shows the cost per step as a function of $N$ up to $N=32$, in \CX{} gates [(a)] and non-Clifford rotations [(b)]. With the factorized coupling, the cost grows nearly linearly with the number of baths, $N-1$: the production configuration requires about 590--700 \CX{} gates and 730--820 rotations per bath and step, and the slow increase of these values reflects the $\log N$ growth of each coupling block. The case $N=3$ lies on the same trend, so the unused basis states of a system register whose dimension is not a power of two add no cost. The number of qubits, Eq.~\eqref{eq:nqubits}, and the number of mid-circuit measurements, one per bath and step, also grow linearly with $N-1$.

Without the factorization, the coupling blocks act on the whole system register and one mode, $\lceil\log_2N\rceil+n_q$ qubits. At $N=3$ and 4 they reach five qubits, still within the exact-synthesis threshold, but the generic synthesis of such blocks is already expensive in rotations, which explains the jump of the unfactorized rotation count between $N=2$ and $N=3$ in Fig.~\ref{fig:scaling}(b). For $N>4$ the blocks exceed the threshold and are implemented with product formulas, which explains the steep increase of the unfactorized \CX{} count between $N=4$ and $N=8$ in Fig.~\ref{fig:scaling}(a). The advantage of the factorization therefore grows with $N$: it reduces the \CX{} count by factors of 1.6, 5.1, 8.9, and 14.2, and the rotation count by factors of 2.3, 1.8, 2.4, and 3.3, at $N=4$, 8, 16, and 32. A single repetition of the dilation instead of two reduces the cost by a nearly constant factor of 1.6--1.8 for all $N$, as expected from a change that affects each dissipative channel in the same way.

Table~\ref{tab:endtoend} summarizes the resources of one trajectory of $n_t=20\,000$ steps. For $N=2$, the production configuration requires 8 logical qubits and $1.6\times10^{9}$ T gates; for $N=32$, a size comparable to the 27 pigments of the LH2 antenna complex, it requires 192 logical qubits and $5.5\times10^{10}$ T gates. These estimates assume baths described by two modes; structured spectral densities would require more modes per bath. The circuit is essentially sequential: the two-qubit depth per step is close to the \CX{} count, because the coupling blocks of all baths act on the system register and all dissipative channels share a single ancilla. With one ancilla per bath, the dilations of different baths would act on disjoint qubits and could run in parallel, which would reduce the depth of the dissipative part of each step, though not the gate count, at the cost of $N-2$ additional qubits. The circuits are far beyond the reach of hardware without error correction~\cite{preskill2018preskill}: one trajectory for $N=2$ already contains about $10^7$ two-qubit gates.

The trajectories are independent and can be distributed over several devices, but their number is fixed by the required statistical precision: the statistical error of $1\times10^5$ trajectories, shown by the bands in Fig.~\ref{fig:indep}(b), is comparable to the error of the fit. On hardware, each shot yields a binary outcome with variance $p(1-p)\leq1/4$, generally larger than that of the estimator used in the emulation, so that more shots are needed for the same precision. The $1\times10^5$ trajectories of the $N=2$ calculation thus amount to about $1.6\times10^{14}$ T gates, and the corresponding calculation for $N=32$ to about $5.5\times10^{15}$.

In fault-tolerant terms, the method is thus demanding in time rather than in space. The number of logical qubits is small compared with quantum-chemistry targets such as FeMoco~\cite{Reiher2017}, which require thousands of logical qubits in the most efficient estimates~\cite{Lee2021}. The total cost, in contrast, is set by the number of time steps and of trajectories, which are fixed by the required discretization and statistical errors, rather than by the compilation of a single step. At fixed precision, this cost can be reduced through higher-order dilated channels, which allow larger time steps for the same discretization error, or through variance-reduction techniques, which reduce the number of trajectories for the same statistical error; otherwise, the rate of magic-state production and the parallel execution of trajectories over several devices are the limiting factors.

\subsection{Comparison with other approaches}
\label{sec:comparison}

As discussed in Secs.~\ref{sec:intro} and~\ref{sec:model:lorentz}, other routes to non-Markovian dynamics on quantum hardware start from generators that are not completely positive, and therefore require a doubled register, postselection, or the readout of amplitudes. Coupled Lindblad pseudomodes avoid all of these requirements: each trajectory uses a single reusable ancilla, keeps both measurement outcomes, and yields the populations as ordinary measurement probabilities, and at fixed time step its cost grows linearly with the simulated time.

The coupled-pseudomode equation can also be simulated with general-purpose Lindbladian algorithms. Methods based on linear combinations of unitaries~\cite{childs2012wiebe,berry2015somma} and quantum signal processing~\cite{Cleve2017,Li2023} require a number of queries that grows linearly with the product of the simulated time and the norm of the Lindbladian, with only a polylogarithmic dependence on the precision, and randomized-dissipation methods~\cite{Kato2024} require a number of segments that grows quadratically with that product. Their favorable dependence on the precision thus comes with a prefactor set by the norm of a Pauli or block encoding, which for bosonic baths is dominated by the jump operators: a truncated ladder operator contains $d\log_2 d$ Pauli strings with coefficients growing as $\sqrt{n}$. Product formulas, such as the circuit of Sec.~\ref{sec:circuit}, have a less favorable dependence on the precision, but their cost is set by the number and width of the blocks, each synthesized from its compact operator form, rather than by the norm of their Pauli decomposition; their splitting error depends on norms of commutators of these blocks, which grow much more slowly with the truncation than the Pauli norm. In terms of scaling, norm-based algorithms are therefore superior in the precision, whereas product formulas have a much smaller prefactor for bosonic baths. The relevant precision is not a free parameter: there is no benefit in simulating the pseudomode model more accurately than the error of the fit, of order $10^{-3}$, or the sampling noise of the trajectories. In this regime, we therefore expect the large norm of the bosonic encoding to outweigh the favorable precision scaling of norm-based algorithms, which would pay off only at much smaller target errors.

For the benchmarks of this work, classical methods are efficient. The cost of HEOM grows with the number of auxiliary density operators, $\binom{\mathcal{N}_{\max}+K}{K}$ for $K$ exponentials in total; for independent baths $K$ grows linearly with $N$, so that the hierarchy grows combinatorially with the number of sites, although free-pole decompositions~\cite{Xu2022,xu2023ankerhold} and tensor-network representations of the hierarchy~\cite{Shi2018,Borrelli2019,Borrelli2021,Takahashi2024,ke2022thoss} reduce this cost in practice. The classical propagation of the coupled-pseudomode model, which we use as a reference, grows exponentially with the number of baths, since the register gains $m\,n_q$ qubits per bath, whereas tensor-network methods such as DAMPF~\cite{Somoza2019,Lorenzoni2024,Lorenzoni2025} avoid this growth when the entanglement between the system and the modes remains moderate. In the quantum circuit, in contrast, the number of qubits and the cost per step grow nearly linearly with the number of baths (Sec.~\ref{sec:resources}), independently of the entanglement generated during the dynamics, and the dissipation damps the numerical errors, so that they do not accumulate over long simulated times (Fig.~\ref{fig:errors}). The quantum approach is therefore expected to become competitive for large systems with strong system--bath coupling and long simulated times, where the entanglement between the system and its environment limits tensor-network methods; establishing the crossover requires a dedicated comparison for specific systems.

In terms of fault-tolerant resources, the method requires few logical qubits (Sec.~\ref{sec:resources}), which makes it a natural candidate for early fault-tolerant devices~\cite{katabarwa2023johnson}, and the non-Clifford cost of a single trajectory, $10^{9}$--$10^{11}$ T gates for $N=2$--32, is of the same order as that of a single phase-estimation run for quantum-chemistry targets such as FeMoco~\cite{Reiher2017}, about $5\times10^{9}$ Toffoli gates in the most efficient estimates~\cite{Lee2021}. Unlike the energy in phase estimation, whose precision is set by the circuit depth, the populations are expectation values that must be estimated from repeated trajectories, whose number grows as the inverse square of the statistical precision. Amplitude estimation could reduce this to the inverse precision, at the cost of a coherent implementation without mid-circuit measurements.
Overall, the cost scales polynomially in all parameters: nearly linearly in the number of baths, linearly in the simulated time at fixed time step, logarithmically in the synthesis precision, and as the inverse square of the statistical precision. These estimates assume logical error rates low enough for each trajectory to run without logical errors; the corresponding overhead in code distance, physical qubits, and magic-state factories depends on the hardware and is not included. Applications to realistic systems will further require more modes per bath for structured spectral densities.

Finally, platforms with native bosonic modes offer a different route. In trapped ions, the pseudomodes can be realized directly as motional modes damped by sympathetic cooling~\cite{Lemmer2018}, programmable simulations of bosonic systems have been demonstrated~\cite{Chen2023,Katz2023}, and models of the same class have already been simulated in this way~\cite{So2024}. This analog formulation is not covered by the present analysis and would require its own: the qubit encoding of the modes and the errors of the splitting and the dilation are replaced by the noise and control errors of the hardware, which are not corrected. Several elements of the construction nevertheless carry over. The coupled-pseudomode fit and its error are unchanged; the momentum basis still removes one bath exactly, now with the Hermitian collective couplings of Sec.~\ref{sec:compilation:momentum}, which reduces the number of modes required; and a rank-one dissipation matrix means that only one mode per bath needs to be cooled. The preferred mode basis, however, might be inverted: cooling a single mode is native, whereas correlated dissipation over several modes is not, so that diagonalizing $K$ and keeping the coherent inter-mode coupling would be natural there, unlike in the digital circuit. Analyzing coupled pseudomodes on such platforms, where the encoding overhead disappears but error correction is unavailable, is an interesting direction for future work.

\section{Conclusions}
\label{sec:conclusions}

We have developed and benchmarked quantum-trajectory circuits for non-Markovian open-system dynamics based on coupled Lindblad pseudomodes. Because the coupled-pseudomode generator is completely positive, each trajectory runs with a single reusable ancilla, without postselection, and yields the populations as ordinary measurement probabilities. Two coupled modes per bath reproduce HEOM within the accuracy of the fit for Drude--Lorentz and Brownian baths, including the low-temperature corrections that independent Lorentzian modes cannot represent, and the optimized dissipation matrix turns out to be effectively of rank one, so that a single dissipative channel per bath suffices.

The cost of the circuits is dominated by the structure of the encoding. Encoding the eigenmodes of the pseudomode Hamiltonian removes the inter-mode Pauli strings and is by far the most effective step, and, for identical baths, the discrete Fourier transform over the bath index removes one bath and turns the remaining couplings into phase waves whose propagators factorize exactly. With these transformations and further compilation steps, the cost per step at $N=2$ is reduced by two orders of magnitude, and the cost grows nearly linearly with the number of baths up to $N=32$, with an advantage of the factorization that grows with $N$. The error budget shows that the Stinespring dilation requires at least a second-order product formula for its implementation, and that all numerical approximations remain several times below the error of the fit, which, together with the sampling noise, sets the accuracy of the method.

In fault-tolerant terms, the method is demanding in time rather than in space: systems of the size of the LH2 antenna complex require about two hundred logical qubits, and a single trajectory requires a non-Clifford cost comparable to that of a phase-estimation run for quantum-chemistry targets, but the time evolution must be sampled over many trajectories. For bosonic baths, scaling arguments suggest that the circuit could be more efficient than general-purpose Lindbladian algorithms in the regime of precision set by the model, since the cost of the latter is dominated by the large norm of the qubit encoding; a quantitative comparison is left for future work. Reducing the number of time steps through higher-order dilated channels and the number of trajectories through variance reduction are therefore the main routes to lower costs.

The momentum-basis construction applies to any system Hamiltonian, provided that the baths are identical and couple through site projectors, as in standard models of the Fenna--Matthews--Olson complex and of ring-shaped antenna complexes, where spatially correlated baths are also exactly circulant. Structured spectral densities will require more modes per bath; since a representation with a single damped mode exists for any number of modes, it can be imposed directly in the fit through a rank-one dissipation matrix, which also reduces the number of fit parameters. The width of each monolithic dilation then grows with the number of modes per bath, whereas each piece of the bilinear dilation still acts on a single mode and the ancilla. Combining these elements with end-to-end fault-tolerant estimates, and comparing them with the cost of HEOM and tensor-network methods for the same systems, would identify the system sizes at which the quantum approach becomes advantageous; we leave this analysis for future work.

\begin{acknowledgments}
We thank Neill Lambert and Paul Menczel (RIKEN) for many insightful discussions on pseudomode methods and their implementation on quantum computers.
This work was supported by MEXT Q-LEAP Program (JPMXS0120319794), and JSPS KAKENHI Grants (JP23H01921). Computations were partly performed using the Research Center for Computational Science, Okazaki, Japan (Project: 26-IMS-C029).
\end{acknowledgments}

\bibliography{refs}

\end{document}